\documentclass[aps,prb,twocolumn,nobalancelastpage,%
footinbib,%
noshowpacs,preprintnumbers,superscriptaddress%
]{revtex4-1}
\usepackage{amsmath,amssymb,amsfonts,fixmath,siunitx}
\usepackage{bm,graphicx,color}
\usepackage[pdftex,colorlinks=true,%allcolors=mygreen,
					 bookmarksnumbered,bookmarksopen=true,bookmarksopenlevel=-1]{hyperref}

\def\dag{^{\dagger}}

\newcommand{\bra}[1]{\langle #1 \vert} 
\newcommand{\ket}[1]{\vert #1 \rangle} 
\newcommand{\ip}[2]{\langle #1 \vert #2 \rangle}  %inner product

\def\a{\alpha}

\def\d{\delta}
\def\D{\Delta}
\def\e{\epsilon}
\def\ve{\varepsilon}

\def\s{\sigma}

\def\w{\omega}

\newcommand{\abs}[1]{\left\vert #1\right\vert}
\newcommand{\appref}[1]{Appendix \ref{#1}}

\def\cp{\citep}

\def\ct{\citet}

\def\eg{e.g.\ }
\def\Eqref{Eq.~\eqref}

\newcommand{\figref}[1]{Fig.~\ref{#1}}
\newcommand{\figsref}[1]{Figs.~\ref{#1}}

\def\ie{i.e.\ }
\newcommand{\ig}[2]{\includegraphics[width=#1\tw,type=#2,ext=.#2,read=.#2]}

\def\lb{\label}

\def\mcal{\mathcal}

\def\non{\nonumber}

\def\oc{Ref.~\onlinecite}

\def\ra{\rightarrow}

\newcommand{\secref}[1]{Sec.~\ref{#1}}

\def\tbf{\textbf}

\renewcommand{\th}[1]{^\tx{#1}}

\def\tit{\textit}

\newcommand{\tl}[1]{_\tx{#1}}

\def\tw{\textwidth}
\def\tx{\text}

\def\wt{\widetilde}

\newcommand{\eq}[1]{\begin{align} #1 \end{align}}
\newcommand{\eqs}[1]{\begin{subequations}\begin{align} #1 \end{align}\end{subequations}}

\def\bar{\begin{array}}
\def\ear{\end{array}}

\def\bce{\begin{center}}
\def\ece{\end{center}}

\def\ben{\begin{enumerate}}
\def\een{\end{enumerate}}

\def\bfi{\begin{figure}[!tbh]\setcapindent{0em}\centering}
\newcommand{\bfiO}[1]{\begin{figure}[#1]\setcapindent{0em}\centering}
\def\efi{\end{figure}}

\def\bit{\begin{itemize}}
\def\eit{\end{itemize}}

\def\bqu{\begin{quote}}
\def\equ{\end{quote}}

\newcommand{\btblO}[2]{\begin{table}[#2]\setcapindent{0em}\centering\begin{minipage}{#1\tw}}
\newcommand{\btbl}[1]{\begin{table}[!tbh]\setcapindent{0em}\centering\begin{minipage}{#1\tw}}
\def\etbl{\end{minipage}\end{table}}

\def\btbr{\centering\vspace{.4em}\begin{tabular}}
\def\etbr{\end{tabular}}

\def\bseq{\begin{subequations}}
\def\eseq{\end{subequations}}

\def\bve{\begin{array}{l}}
\def\eve{\end{array}{l}}

\newcommand{\igS}[3]{%
\hspace{-.025\tw}\begin{minipage}[t]{#1\tw}\vspace{0em}\includegraphics[width=\linewidth,type=#2,ext=.#2,read=.#2]{#3}\end{minipage}%
\hspace{.03\tw}\begin{minipage}[t]{1\tw-#1\tw-.05\tw}}

\newcommand{\lbS}[1]{\lb{#1}\end{minipage}}

\makeatletter
  \newcommand{\myhref}[2]{\hyper@linkurl{#2}{#1}}
\makeatother

\makeatletter
\newcommand*\if@single[3]{%
  \setbox0\hbox{${\mathaccent"0362{#1}}^H$}%
  \setbox2\hbox{${\mathaccent"0362{\kern0pt#1}}^H$}%
  \ifdim\ht0=\ht2 #3\else #2\fi
  }
\newcommand*\rel@kern[1]{\kern#1\dimexpr\macc@kerna}
\newcommand*\widebar[1]{\@ifnextchar^{{\wide@bar{#1}{0}}}{\wide@bar{#1}{1}}}
\newcommand*\wide@bar[2]{\if@single{#1}{\wide@bar@{#1}{#2}{1}}{\wide@bar@{#1}{#2}{2}}}
\newcommand*\wide@bar@[3]{%
  \begingroup
  \def\mathaccent##1##2{%
    \if#32 \let\macc@nucleus\first@char \fi
    \setbox\z@\hbox{$\macc@style{\macc@nucleus}_{}$}%
    \setbox\tw@\hbox{$\macc@style{\macc@nucleus}{}_{}$}%
    \dimen@\wd\tw@
    \advance\dimen@-\wd\z@
    \divide\dimen@ 3
    \@tempdima\wd\tw@
    \advance\@tempdima-\scriptspace
    \divide\@tempdima 10
    \advance\dimen@-\@tempdima
    \ifdim\dimen@>\z@ \dimen@0pt\fi
    \rel@kern{0.6}\kern-\dimen@
    \if#31
      \overline{\rel@kern{-0.6}\kern\dimen@\macc@nucleus\rel@kern{0.4}\kern\dimen@}%
      \advance\dimen@0.4\dimexpr\macc@kerna
      \let\final@kern#2%
      \ifdim\dimen@<\z@ \let\final@kern1\fi
      \if\final@kern1 \kern-\dimen@\fi
    \else
      \overline{\rel@kern{-0.6}\kern\dimen@#1}%
    \fi
  }%
  \macc@depth\@ne
  \let\math@bgroup\@empty \let\math@egroup\macc@set@skewchar
  \mathsurround\z@ \frozen@everymath{\mathgroup\macc@group\relax}%
  \macc@set@skewchar\relax
  \let\mathaccentV\macc@nested@a
  \if#31
    \macc@nested@a\relax111{#1}%
  \else
    \def\gobble@till@marker##1\endmarker{}%
    \futurelet\first@char\gobble@till@marker#1\endmarker
    \ifcat\noexpand\first@char A\else
      \def\first@char{}%
    \fi
    \macc@nested@a\relax111{\first@char}%
  \fi
  \endgroup
}
\makeatother

\def\figdirCheb{./}
\def\numAA{0.46}
\def\numA{0.4}
\def\numB{0.22}

\begin{document}

\title{Spectral functions and time evolution 
from the Chebyshev recursion}

\author{F. Alexander Wolf}
\affiliation{
Department of Physics,
Arnold Sommerfeld Center for Theoretical Physics,
LMU Munich,
Theresienstrasse 37,
80333 M\"unchen, Germany}
\author{Jorge A. Justiniano}
\affiliation{
Department of Physics,
Arnold Sommerfeld Center for Theoretical Physics,
LMU Munich,
Theresienstrasse 37,
80333 M\"unchen, Germany}
\author{Ian P. McCulloch}
\affiliation{
Centre for Engineered Quantum Systems,
School of Physical Sciences,
The University of Queensland,
Brisbane, Queensland 4072, Australia}
\author{Ulrich Schollw\"ock}
\affiliation{
Department of Physics,
Arnold Sommerfeld Center for Theoretical Physics,
LMU Munich,
Theresienstrasse 37,
80333 M\"unchen, Germany}

\date{\today}

\begin{abstract}
We link linear prediction of Chebyshev and Fourier expansions 
to analytic continuation.
We push the resolution in the Chebyshev-based computation of $T=0$ many-body
spectral functions to a much higher precision by 
deriving a modified Chebyshev series expansion that allows
to reduce the expansion order by a factor $\sim\frac{1}{6}$. 
We show that in a certain limit the Chebyshev technique   
becomes equivalent to computing spectral functions
via time evolution and subsequent Fourier transform. 
This introduces a novel recursive time evolution algorithm
that instead of the group operator $e^{-iHt}$
only involves the action of the generator $H$. 
For quantum impurity problems, we introduce
an adapted discretization scheme for the bath spectral function.
We discuss the relevance of these results for 
matrix product state (MPS) based DMRG-type algorithms,
and their use within dynamical mean-field theory (DMFT).
We present strong evidence that the Chebyshev recursion extracts
less spectral information from $H$ than time 
evolution algorithms when fixing a given amount of created entanglement.
\end{abstract}

%\pacs{71.27.+a, 71.10.Fd, 05.70.Ln}

\maketitle

%\tableofcontents

%
\section{Introduction}

Expanding the spectral density $A(\w)$
of an operator $H$ in the monomes $\w^n$ via the moments
$$
\mu_n\th{mon}=\int d\w\,A(\w) \w^n,
$$ 
is a tool that originates in the early days of 
quantum mechanics.\cp{weisse06}
Computing these moments iteratively though
is numerically unstable\cp{gautschi67,gautschi68} and one
replaced expansions in $\w^n$ by expansions
in such polynomials $p_n(\w)$ of degree $n$ 
that can be stably computed.\cp{gautschi70,sack71}
A prominent example for $p_n(\w)$ are Chebyshev polynomials,
whose associated three-term recursion is stable as it does not
admit a so-called \tit{minimal}\cp{gautschi67} solution.

After the development of stable recursions
the next step in the mid 1990s was 
the introduction of kernels that damp 
the erroneous \tit{Gibbs oscillations} 
of truncated polynomial expansions
of discontinuous functions,\cp{silver94,wang94,wang94i}
which lead to the \tit{kernel polynomial approximation}. 
It deals with redefined series expansions 
that represent the convolution
of the expanded function with a \tit{broadening kernel}, like a Gaussian 
or Lorentzian. This technique has been reviewed in \oc{weisse06} 
and more recently in \oc{lin13} from a numerical linear
algebra perspective.

In this paper, we drop the idea of such \tit{broadening} kernels in frequency space 
or the equivalent \tit{damping} or \tit{windowing} kernels  
in the associated Fourier or Chebyshev expansions. 
Instead, we employ the fundamentally different technique 
of \tit{linear prediction}.\cp{numrec07} 
Linear prediction is a \tit{linear} recursive reformulation (\appref{secLP})
of the \tit{non-linear} problem to fit the surrogate function 
\eq{ \label{eqLPansatz}
g(t) = \sum_i \alpha_i e^{i \w_i t}, \quad \a_i,\w_i \in \mathbb{C}, ~ t \in \mathbb{R},
}
to given numerical data $\{t_n,g_n\}$. Due to linearity, 
linear prediction is able to treat superpositions 
of hundreds of terms, and by that reliably 
extracts much information about an underlying function 
from its \tit{local} 
knowledge $\{t_n,g_n\}$. 
In order for this to be meaningful, the 
underlying function, \eg
a Green's function, 
must be compatible with \eqref{eqLPansatz}.

In particular, we note that \eqref{eqLPansatz} can serve as
an ansatz for \tit{analytic continuation} of a zero-temperature Green's function
\eq{ \label{green}
G(t) = -i \bra{\psi_0} e^{-i(H-E_0)t} \ket{\psi_0}, 
}
where $\ket{\psi_0}$ is a single-particle excitation of the
groundstate $\ket{E_0}$ of $H$, 
for example the creation of a 
fermion $\ket{\psi_0} = c\dag\ket{E_0}$. Note that 
in the case of fermions, \eqref{green} describes 
only the $t>0$ contribution (then usually more precisely denoted $G^>(t)$)  
of the \tit{full} fermionic Green's function.   
$G(t)$ is analytic everywhere in the complex plane except for $t \ra i\infty$
and thereby allows for an analytic continuation of $G(t)$  
from a local description $\{t_n,G(t_n)\}$ to the domain $[t_0,\infty)$. 
This analytic continuation is highly different from 
the ill-conditioned problem of continuing the frequency-space  
represented Green's function from a domain in the complex plane 
(\eg the imaginary-frequency axis or a parallel
of the real-frequency axis) 
to the real-frequency axis, where 
the frequency-space Green's function has poles. 

In the context of Green's functions, 
linear prediction has for the first time been used to 
extrapolate the time evolution of  
the \tit{spin structure factor}
in the one dimensional Heisenberg model.\cp{white08,barthel09}  
While for the spin-$1$ model it was clear that 
the ansatz \eqref{eqLPansatz} is justified as 
the time evolution is dominated by 
a small number of \tit{magnons} 
whose excitation energies  
correspond directly to the 
frequencies $\w_i$ in \eqref{eqLPansatz},\cp{white08}
this was not the case for the spin-$\frac{1}{2}$ model.\cp{barthel09}  In the latter,
\tit{spinons} dominate which lead to an (infinitely) high number of poles
on the real-frequency axis, 
and the \tit{direct} correspondence of pole energies and 
frequencies $\w_i$ in \eqref{eqLPansatz} is lost.
Still the ansatz works\cp{barthel09} in an approximate sense by extracting 
effective frequencies.  

For the computation of spectral functions, 
the use of linear prediction for the time evolution 
of Green's functions provides 
a highly attractive alternative approach to the usual 
\tit{damping} or \tit{windowing} in real-time or 
\tit{broadening} in frequency space: 
an approach that \tit{enhances} resolution in frequency space. 
Up to now, it is not entirely clear in which cases this 
is controlled. On the other hand,  
the approach of \tit{damping} the  
truncated series expansion cannot be considered controlled, too:  
Although a broadened function $f_\eta(\w)$, 
which is for Gaussian broadening given by 
$f_\eta(\w)=\frac{1}{\eta\sqrt{2\pi}}\int d\w' e^{-(\w'-\w)^2/2\eta^2}f(\w')$, 
converges uniformly to the 
underlying original function $f(\w)$ for $\eta\ra 0$, extraction of information 
(\tit{deconvolution}) from $f_\eta(\w)$ about $f(\w)$ is uncontrolled as it 
corresponds to the problem of analytic continuation from a domain in the complex 
plane to the real axis.

Recently, \oc{ganahl14} suggested to extrapolate the Chebyshev expansion
of a spectral function using linear prediction, 
albeit only justified by the empirical success. In the remainder of this introduction,
we place these results in the context of the preceding discussion, and by 
that put this approach on more firm grounds.

\subsection{Chebyshev and Fourier transformation basics}
\label{secChe}

The Chebyshev polynomials of the first kind
\eq{
  T_{n}(x) = \cos\left(n \arccos(x)\right)   \label{Tn}
}
can be generated by the recursion
\eq{ \label{recScalar}
  T_{n}(x) = 2xT_{n-1}(x) - T_{n-2}(x), \quad
  T_{1} = x, \quad T_{0} = 1,
}
which is numerically stable if $\abs{x} \leq 1$. Chebyshev 
polynomials are orthonormal with respect to the weighted 
inner product
\eqs{
 \int_{-1}^{1} dx\, w_n(x) T_m(x) T_n(x) & = \d_{nm},  \label{eqOrtho} \\
 w_n(x) & = \frac{2-\d_{n0}}{\pi \sqrt{1-x^2}}.     \label{eqMeasure}
}
Any integrable function $f(x)\vert_{x\in[-1,1]}$ can be expanded in $T_n(x)$
\begin{subequations}
\label{chebExp}
\eq{
  f(x) &=
  \sum_{n=0}^{\infty} w_n(x) \mu_{n} T_{n}(x), \label{chebExp1} \\
  \mu_n &= \int_{-1}^{1} dx\, T_n(x) f(x),   \label{eqScalarProd}
}
\end{subequations}
where the definition of the so-called \tit{Chebyshev moments} $\mu_n$ via
the \tit{non-weighted} inner product \eqref{eqScalarProd} 
follows when applying $\int_{-1}^{1}dx\, T_m(x) \dots$
to both sides of  \eqref{chebExp1}.

Analogously, any integrable function $f(\w)\vert_{\w\in[-\frac{a}{2},\frac{a}{2}]}$, 
where $a\in\mathbb{R}$, can be expanded in a Fourier series
\eqs{
f(\w) & = \frac{1}{2a\pi}  \sum_{n=-\infty}^{\infty} e^{i\w t_n} f(t_n), \\
f(t_n)  & =  \int_{-a/2}^{a/2} d\w\, e^{-i\w t_n} f(\w), \quad t_n=\frac{n}{a},
}
which represents a \tit{Fourier transform} for $a\ra\infty$.

\subsection{Expansion of a spectral function}
\label{secChe}

Consider now the expansion of the spectral function $A(\w)$
of a Hamilton operator $H$
with respect to a reference energy $E\tl{ref}$ and a state $\ket{\psi_0}$ 
as in \eqref{green}
\eq{ \label{specFunc}
A(\w) 
& = \bra{\psi_0} \d(\w - (H-E\tl{ref})) \ket{\psi_0}.
}
The spectral function is related to 
the Green's function of \eqref{green} 
via its Fourier transform: 
$A(\w)= -\frac{1}{\pi}\tx{Im}\,G(\w+i0^+)$.

The coefficients of the Fourier expansion can be computed 
by inserting an identity  of eigenstates $\sum_i\ket{E_i}\bra{E_i}$ 
in the integral over the delta function $\delta(\w - (H-E\tl{ref}))$ 
\begin{subequations}
\label{fourCoeff}
\eq{
f(t_n) & =  \int_{-a/2}^{a/2} d\w\, e^{-i\w t_n} A(\w)  
	 =  \ip{\psi_0}{\psi(t_n)}, \label{ft} \\
\ket{\psi(t_n)}  & = e^{-i(H-E\tl{ref})t_n} \ket{\psi_0}, \quad t_n=\frac{n}{a}.   \label{psit}
}
\end{subequations}
In order for \eqref{fourCoeff} to 
hold true, 
$a$ must be chosen large enough for that the support of $A(\w)$ 
is contained in $[-\frac{a}{2},\frac{a}{2}]$. 
A sufficient condition for that is  $\tx{spec}(H-E\tl{ref}) \subset [-\frac{a}{2},\frac{a}{2}]$, 
which is possible as we consider operators $H$ with bounded spectra. 
\Eqref{psit} makes it obvious that $a$ has the 
meaning of an inverse time step.

To compute the coefficients for the Chebyshev expansion, we need 
to consider a spectral function whose support is contained in $[-1,1]$. 
For this, introduce a rescaled and shifted version of 
$H$ with appropriately chosen constants $a$ and $b$ 
\eq{
  \mathcal{H}_{a,b} = \frac{H-E\tl{ref}}{a} + b, \quad x = \frac{\w}{a} + b,       \label{eqShiftScale}
}
where $a$ can again be considered an 
``inverse time step" and $\mathcal{H}$ is dimensionless.
Note that in \oc{wolf14}, the definition of $b$ differed from the one here by a factor $a$.
Then
\eq{ 
  \mathcal{A}_{a,b}(x) = \bra{\psi_0} \delta(x - \mathcal{H}_{a,b}) \ket{\psi_0}  \label{eqAscaled} 
}
yields the original spectral function via $A(\w) =  \frac{1}{a} \mathcal{A}\big(\tfrac{\w}{a}+b\big)$,
where we omitted to specify the indices $a,b$, as in most of the rest of this paper.
The Chebyshev moments for $\mathcal{A}(x)$ can be computed 
analogously to the Fourier coefficients \eqref{fourCoeff}
\begin{subequations}
\label{eqMuH}
\eq{ 
  \mu_n  & = \int_{-1}^{1} dx\,\mathcal{A}(x) T_n(x) = \ip{\psi_0}{\psi_n},  \label{mun}\\
  \ket{\psi_n} & = T_n(\mathcal{H}) \ket{\psi_0}.   \label{tn}
}
\end{subequations}

Inserting the recursive definition \eqref{recScalar} of $T_n(\mathcal{H})$ in the definition \eqref{tn} of $\ket{\psi_n}$,
one obtains a practical calculation scheme for the power series expansion of $T_n(\mathcal{H})$,
and by that for the \tit{Chebyshev states} $\ket{\psi_n}$ in \eqref{eqMuH}:  
\eq{ 
  \ket{\psi_{n}}  = 2 \mathcal{H} \ket{\psi_{n-1}} - \ket{\psi_{n-2}}, \quad 
  \ket{\psi_{1}} = \mathcal{H} \ket{\psi_{0}}.   \label{rec}
}

\subsection{Analytic continuation}

A comparison of \eqref{psit} and \eqref{tn} clarify why linear prediction
is an equally justified approach for a Chebyshev as for a Fourier expansion.

Rewriting the ``evolution operators" that appear in \eqref{psit} and \eqref{tn} as
\eq{
 \exp(-in\mcal{H}_{a,b=0})     \qquad \tx{and} \qquad  \cos(n \arccos(\mcal{H}_{a,b}))
}
makes it clear that we deal with \tit{analytic} functions of $n$
if we consider $n$ as a continuous complex variable. 
By \eqref{ft} and \eqref{mun}, 
this makes the Fourier and the Chebyshev coefficients 
$f(t_n)$ and $\mu_n$ analytic functions of $n$, too. 
In addition, $\tx{Re}\,  \exp(-in\mcal{H}_{a,b}) = \cos(n\mcal{H}_{a,b})$ shows
that the real part of the functional dependence of the time evolution operator 
on $n$ is the same as for the ``Chebyshev evolution operator",
when neglecting a redefinition of oscillations frequencies. As this redefinition can be
accounted for by the fitting procedure, the particular 
form of the surrogate function $g(t)$ in \eqref{eqLPansatz} is equally suited 
to analytically continue both types of expansions. A 
fundamental theorem from complex analysis then tells us that 
if linear prediction provides us with a function  $g(t)$ 
that \tit{locally agrees} with $f(t_n)$ or $\mu_n$, we know that this 
function globally agrees with $f(t_n)$ or $\mu_n$. 
Of course, in practice these arguments are to be taken with care, 
as we will never numerically find a function $g(t_n)$
that \tit{agrees exactly} with the local data $\{g_n,t_n\}$.

\subsection{Outline of the paper}

We will first study the convergence properties of the Chebyshev
expansion of discontinuous (spectral) functions in the 
thermodynamic limit. This allows to derive a new scheme
for a Chebyshev series defintion that leads to exponential 
convergence and allows to reduce expansion orders
in practical calculations by a factor $\sim\frac{1}{6}$ (\secref{secTL}). 
We then apply these results to the computation
of spectral functions for finite systems (\secref{secFS}),
and discuss the relevance for 
matrix product state (MPS) based computations (\secref{secMPS}).
After that, we describe the approximate 
equivalence of the Chebyshev recursion to 
time evolution and show how this leads
to a novel time evolution algorithm (\secref{secTimeEvol}).  
Finally, we conclude the paper (\secref{secCon}).

\section{Spectral functions in the thermodynamic limit}
\label{secTL}

A spectral function for a system of  
finite size $L$ has a \tit{finite-size peak} structure 
due to an agglomeration of eigenvalues that is not present in the 
thermodynamic limit.
In a weakly interacting system, this agglomeration happens  
around the positions of the eigenvalues 
of the corresponding noninteracting 
(single-particle) system.
This argument gives us the best, though still very rough, 
estimate $W\tl{single}/L$ for the spacing of finite-size peaks, 
where $W\tl{single}$ is the single-particle bandwidth. 
At a much smaller spacing than that, 
spectral functions have an underlying 
delta peak structure, as is obvious from definition \eqref{specFunc},
which can be rewritten as
\eq{ \label{specFunc1}
A(\w) & = \sum_i W_i\,\d(\w - (E_i-E\tl{ref})),
}
with weights $W_i = \abs{\ip{\psi_0}{E_i}}^2$. 
The delta peak structure
merges to a (section-wise) smooth function
only in the thermodynamic limit.

Expanding the spectral function
of a finite-size system in orthogonal polynomials is a very efficient way 
to \tit{not} resolve either finite-size peaks 
or the delta peak structure, but to extract only the 
smooth function of the thermodynamic limit, 
as \eg discussed in \oc{wolf14}. It is this function of the
thermodynamic limit that we are interested in, and 
for which we start our discussion.

\subsection{Discontinuity of spectral functions}

The state $\ket{\psi_0}$ and 
the energy $E\tl{ref}$ in \eqref{specFunc1} 
are generally associated to the ground state of 
a certain symmetry sector $N$ of $H$, 
which for fermions is typically a particle number. 
The reference energy for  
$\ket{\psi_0} = c\dag\ket{E_0}$ then is the Fermi energy,
which is the ground state energy 
$E\tl{ref} = E_0^{N-1}$ of the contiguous symmetry sector of $\ket{\psi_0}$ 
(or $E\tl{ref} = E_0^{N+1}$ for a hole excitation). 
The weights $W_i = \abs{\ip{\psi_0}{E_i}}^2$
in the spectral function \eqref{specFunc1} can be non-zero 
only for eigenstates $\ket{E_i}$ and eigenvalues $E_i$ from the sector $N$.
The particular meaning of $E\tl{ref}$ as a ground state energy
then implies that even if the global spectral function 
\eq{
A\tl{global}(\w) = \sum_i \d(\w - (E_i-E\tl{ref}))
}
is smooth, the weights $W_i$ generally introduce
a discontinuity at $\w=0$
(we use the term \tit{global}
here, as $\ket{\psi_0}$ usually is a \tit{local} excitation 
associated with a certain quantum number).

\subsection{Convergence of Chebyshev series expansions}

The convergence of the Chebyshev moments 
$\mu_n \ra 0$ of a function $f(x)$ \tit{in the limit} $n\ra\infty$
can be characterized by the degree of differentiability of $f(x)$,
similar to a Fourier expansion.\cp{boyd01} 
Let $k$ denote the highest integer for which the $k$th derivative 
of $f(x)$ is integrable: If $f(x)$ is smooth ($k=\infty$), the envelope of $\mu_n$  
converges exponentially to zero with respect to $n$; 
if $f(x)$ is a step function ($k=1$), the envelope converges algebraically with $\frac{1}{n}$; 
and if $f(x)$ is a delta function, the envelope remains constant. 
In general, the order of convergence is \tit{at least} $\frac{1}{n^{k}}$.
Although in \oc{boyd01}, this is stated for moments computed 
with the \tit{weighted} inner product \eqref{eqMeasure}, it also holds for moments 
computed using \eqref{eqScalarProd} (see \appref{secConv}). 
In practice, we are not interested 
in the limit $n\ra\infty$, but rather in intermediate values of $n$: 
but also here, the degree of differentiability of $A(\w)$ 
helps us to learn something about the convergence of $\mu_n$.

\begin{figure}
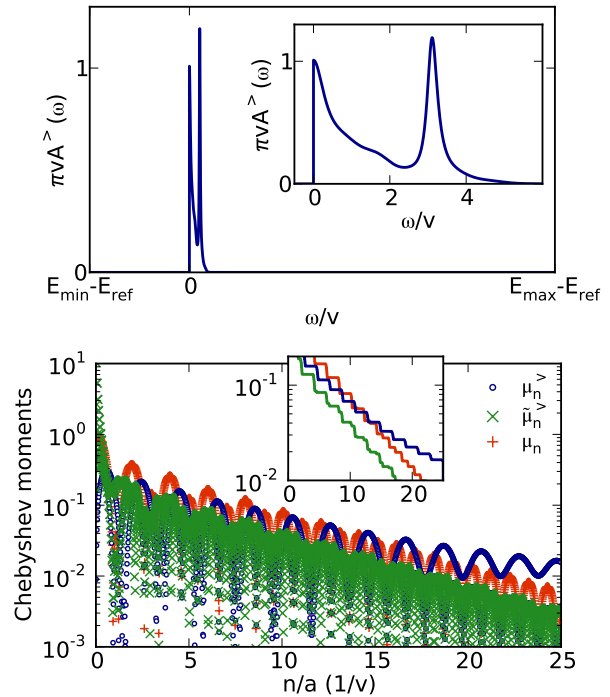

\ig{\numAA}{pdf}{\figdirCheb fig_specExample}
\ig{\numAA}{pdf}{\figdirCheb fig_mu_cmpTL}
\caption{(Color online) 
\tbf{Top}:
Typical example of a discontinuous spectral function
due to the restriction to a given symmetry sector.
In this case, this is the particle contribution of the spectral function 
of the single-impurity Anderson model (SIAM)  
with semi-elliptic bath density of states of half-bandwidth $2v$
and interaction $U/v=4$,\cp{raas04} taken from \oc{wolf14}.  
The spectral function is given by \eqref{specFunc} 
using $\ket{\psi_0} = c\dag\ket{E_0}$,
where $\ket{E_0}$ denotes the
half-filled ground state and $E\tl{ref}$ is the Fermi energy $E_0$. 
Here, only the shape of the scalar function is of importance, therefore
we postpone the model definition to \eqref{Hsiam}. 
The same spectral function is obtained for the local
density of states of the first site
for spinless fermions hopping on a semi-infinite chain
with tunneling $v$ and and an interaction
of $U/v=4$ that acts only on the first site.
\tbf{Bottom}:
Comparison of convergence of 
the Chebyshev moments  
of $A^>(\w)$ with its redefinitions  
$\wt A^>(\w)$ and $A(\w)$, giving rise
to moments $\mu_n^>$, $\wt\mu_n^>$ and $\mu_n$,
respectively. 
The full spectral function \eqref{eqFull} for this example is    
$A(\w) = A^>(\w) + A^>(-\w)$ as here, 
$A^<(\w)=A^>(\w)$ due to particle-hole symmetry. 
All of this is for the setup $b=0$ using 
a rescaling of $a=100v$ in \eqref{eqShiftScale}. 
}
\label{figSpecExample}
\end{figure}

Consider a typical discontinuous
spectral function $A^>(\w)$ as shown 
in the top panel of  \figref{figSpecExample}.
Its corresponding Chebyshev moments $\mu_n^>$ are 
computed by numerically integrating \eqref{eqScalarProd} and  
shown in the bottom panel of \figref{figSpecExample} 
as blue circles. The blue line in the inset 
shows the envelope of $\mu_n^>$, which  
evidently decreases algebraically to zero.

Now note that continuity of $A^>(\w)$ at $\w=0$ 
can easily be restored by defining
\eq{\label{eqSub}
\wt A^>(\w) = A^>(\w)-A^>(0).
}
The green crosses (lines) in the bottom panel 
of \figref{figSpecExample} show that the 
Chebyshev moments $\wt \mu_n^>$ of $\wt A^>(\w)$ converge exponentially
for the values of $n$ considered in the plot,
i.e., \tit{qualitatively} differently than $\mu_n^>$.   
This is observed although $\wt A^>(\w)$ 
is not smooth, but only once  
differentiable (kink in first derivative at $\w=0$).

While the construction of $\wt A^>(\w)$ is completely general, 
for the particular case of a fermionic 
spectral function, another way of constructing a continuous 
function from $A^>(\omega)$ has been favored: 
In the appendix of \oc{holzner11}, it was mentioned  
that the Chebyshev expansion of the \tit{full} 
spectral function 
\eq{  
A(\w) & = A^>(\w) + A^<(-\w) , \label{eqFull} \\
A^\gtrless(\w) & = \bra{\psi_0^\gtrless} \, \d(\w - (H - E_0)) \ket{\psi_0^\gtrless}, \non
}
obtained by summing over particle 
($>$) and hole ($<$) contributions,  
is much better suited for a Chebyshev expansion than $A^{\gtrless}(\w)$,
as it lacks the discontinuity. 
In \oc{ganahl14} it was then pointed out that 
the \tit{full} $A(\w)$ is \tit{smooth}  
and therefore, Chebyshev moments should decrease
exponentially, which  
would allow to use linear prediction. 
In general, it is \tit{not} true that $A(\w)$ is smooth, 
due to the possibility of \tit{van Hove} singularities, as appear
\eg for the $U=0$ case of the spectral function of the single impurity Anderson model (SIAM) 
(see \appref{secChebExamples}, \figref{figU0}).
Still, $A(\w)$ is likely to be smooth, and for the present example,
it is. 
The bottom panel of \figref{figSpecExample} 
therefore shows that the Chebyshev moments $\mu_n$ for $A(\w)$ 
decrease at the same exponential rate as the moments 
$\wt \mu_n^>$ of $\wt A^>(\w)$.

The statements about the qualitatively 
different convergence behaviors of $A^>(\w)$, $\wt A^>(\w)$ 
and $A(\w)$ are confirmed for further typical examples
in \appref{secChebExamples}.

\subsection{Comparison of setups $b=0$ and $b\simeq -1$}

In \oc{wolf14}, we pointed out that the choice $b=0$ in \eqref{eqShiftScale} 
is computationally much less efficient than the choice  $b\simeq-1$
(called ``$b\simeq-a$" in \oc{wolf14}). 
Whereas constructing a Chebyshev 
expansion of the \tit{full} spectral function $A(\w)$
\tit{requires} choosing $b=0$, this is not the
case for $\wt A(\w)$. For $\wt A(\w)$, we can therefore
use the exponential rate of convergence to quantify 
the amount of spectral information   
that the Chebyshev recursion extracts
from $H$ in the setups $b=0$ and $b\simeq-1$,
and by that understand the observations of \oc{wolf14} quantitatively.

The key observation to make is that the integral 
\eq{ \label{muPlain}
\mu_n^> 
= \int_{-1}^{1} dx\, \mathcal{A}_{a,b}^>(x) T_n(x)
}
extracts a highly different amount 
of information about the 
structure of $A^>(\w)$ depending
on how  $a$ and $b$ in \eqref{eqShiftScale} are chosen
when generating  $\mathcal{A}_{a,b}^>(x)$. 

Throughout the whole paper, we keep $a=100v$ fixed to guarantee the
numerical stability of the Chebyshev recursion
for the typical system sizes of around $L\geq80$ 
that are large enough to display ``thermodynamic limit behavior". 
If we chose $a$ smaller, we could only stably compute ``small" systems 
or we would have to resort to the technique of 
\tit{energy truncation}, which is strongly prone to errors.\cp{holzner11}
Furthermore, in the MPS context, it is important to compare only computations
in which $a$ is kept constant: 
constant $a$ means constant effective hopping 
energies $\frac{v}{a}$ in $\mcal{H}$, 
and by that a constant amount of entanglement production in 
a single iteration step of \eqref{rec}.   
The parameter $b$, by contrast, can be chosen freely 
without affecting the numerical stability, 
and in principle, without affecting 
entanglement production in MPS computations.

\begin{figure}
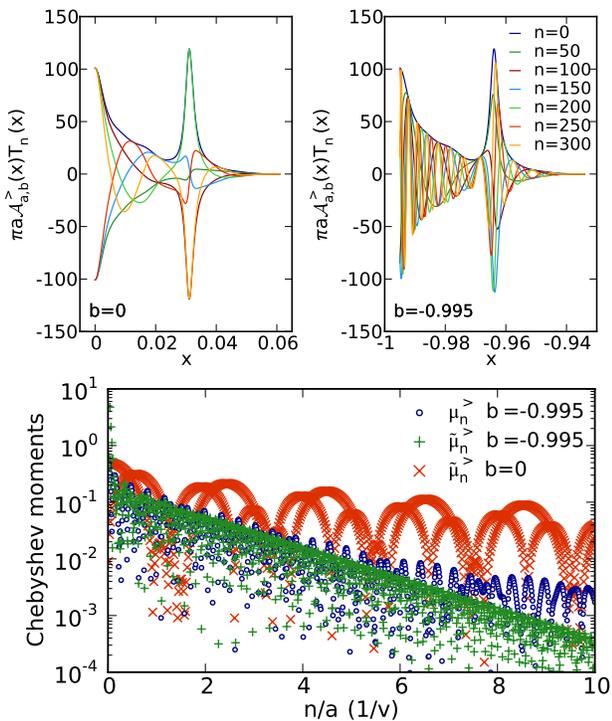

\ig{\numB}{pdf}{\figdirCheb fig_integrand_ofs1.0}
\ig{\numB}{pdf}{\figdirCheb fig_integrand_ofs0.005}
\ig{\numAA}{pdf}{\figdirCheb fig_mu_cmpTLo}
\caption{(Color online) 
\tbf{Top}: 
Integrand for computation of moments for the
spectral function shown in \figref{figSpecExample}. 
in the two setups 
$b=0$ (\tbf{left})
and
$b=-0.995$  (\tbf{right}). 
\tbf{Bottom}: 
Comparison of convergence of 
moments computed with the integrands
shown in the top panels.
In all of that, $a=100v$.
}
\label{figIntegrand}
\end{figure}

The top panels of \figref{figIntegrand}
show the convolution of $\mathcal{A}_{a,b}^>(x)$
with Chebyshev polynomials $T_n(x)$ of different
degree $n$ for the two setups $b=0$ and $b=-0.995\simeq-1$. 
The highly increased oscillation frequency 
that is evident in the setup $b=-0.995$ can be understood 
by looking at the natural stretching of the frequency scale
of Chebyshev polynomials close to the boundaries of $[-1,1]$. 
Expressing the integral \eqref{muPlain} 
by substituting $x=\cos\theta$
\eq{ \label{intStretch}
\mu_n^> 
= -\int_{\pi}^{0} d\theta\, \mathcal{A}_{a,b}^>(\cos\theta) \cos(n\theta) \sin\theta,
}
one arrives at a convolution with the
regularly oscillating $\cos(n\theta)$. 
Consider now the interval of width $0.05$ on $[-1,1]$,
which corresponds to the (single-particle) support of $\mathcal{A}_{a,b}^>(x)$ 
in the example of \figref{figIntegrand}. 
By computing the integral widths under the map $x=\cos \theta$, one learns that placing 
the support in the ``boundary region'' 
$[-0.995,-0.95]$, as results for $b=-0.995$, 
increases the resolution by a factor $\sim 6.4$
compared to placing it in the ``center region" $[0,0.05]$, 
as results for $b=0$.
These effects are well known boundary effects of the Chebyshev
polynomials that are exploited also in the 
solution of differential equations.\cp{boyd01} 

The bottom panel of \figref{figIntegrand}  
shows the Chebyshev moments
obtained in the $b=-0.995$ setup, 
for $A^>(\w)$ (blue circles) and for $\wt A^>(\w)$ (green pluses). 
As mentioned before, 
for this setup, 
no Chebyshev expansion of the \tit{full}
spectral function $A(\w)$ is possible. 
Instead, we compare the $b=-0.995$ results
to the $b=0$ results, depicted as red crosses. 
It is evident that the Chebyshev expansion in the $b= -0.995$
setup converges much faster than the one in the $b=0$ setup.
After $n=1000$ iterations, the magnitude differs by more than 100. 

This difference directly appears in the error of the Chebyshev series,
as stated by the following general rule: 
the order of the error $\varepsilon$ of a Chebyshev 
(or Fourier) series representation of a function that is truncated at $n=N$ 
can be estimated by (see \oc[Chap.~2.12]{boyd01})
\begin{subequations}
\label{convEstim}
\eq{
\varepsilon & = \mcal{O}(\mu_N) \quad\tx{ if $\mu_n$ converges exponentially}, \\
\varepsilon & = \mcal{O}(N \mu_N) \quad\tx{ if $\mu_n$ converges algebraically}.
}
\end{subequations}

\subsection{Linear prediction for the Chebyshev expansion}

The main motivation for studying the convergence  
of different Chebyshev expansions in the previous sections 
lies in the possibility to extrapolate
exponentially decreasing sequences 
with \tit{linear prediction}. 
As discussed in the introduction, 
the latter allows an extremely high gain in resolution,
if its application is justified. For details on
linear prediction, see \appref{secLP}.

In what follows, we compare
the known approach of using linear prediction
for the Chebyshev expansion of $A(\w)$, as suggested 
in \oc{ganahl14},  
with the new approach of extrapolating the  
Chebyshev expansion of $\wt A^>(\omega)$.

We first compute the Chebyshev moments 
of the step function that has the discontinuity 
of $A^>(\w)$ at $\w=0$, which transforms to $x=-b$
for $\mcal{A}^>(x)$, as 
\eq{
\mu_n^{\tl{step}} 
& = \int_{-b}^{1} dx\,T_n(x) \\
& = \frac{1}{2} \Big( \frac{\cos[(n+1)\arccos x]}{n+1} \non \\
& \qquad~    - \frac{\cos[(n-1)\arccos x]}{n-1}\Big) \Big\vert_{-b}^1 \non .
}
The Chebyshev moments of $\wt A^>(x)$ are then given by
\eq{
\wt \mu_n^>
= \mu_n^> - A^>(0) \mu_n^{\tl{step}}
}
and are accessible by linear prediction,
as they decrease exponentially.

The core problem in this new approach is
that the value $A^>(0)$ of the spectral function is,  in general, 
unknown \tit{prior} to linear prediction. But 
it fulfills the following self-consistency problem, which can be iteratively solved:
Choosing a start value $A_0^>(0)$ for $A^>(0)$, we compute 
$\wt\mu_n$, extrapolate the sequence up to convergence, 
and then use the extrapolated sequence to  
reconstruct $A^>(\w)$,
which provides us with a new value $A_1^>(0)$. We repeat
the procedure until the new and the old version $A_{i}^>(0)$ 
and $A_{i+1}^>(0)$ agree. This procedure is found
to converge stably and quickly for all examples studied
(see also \appref{secChebExamples}).

\begin{figure}
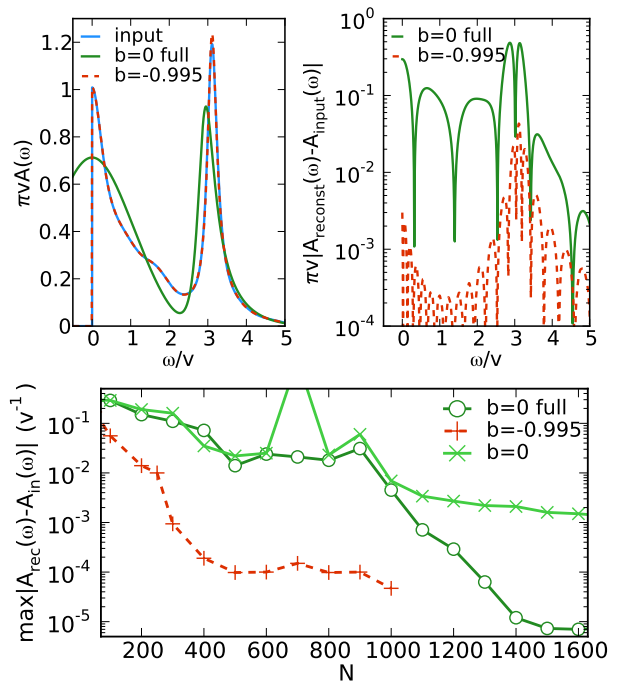

\ig{\numB}{pdf}{\figdirCheb fig_U4_rho_TL}
\ig{\numB}{pdf}{\figdirCheb fig_U4_rhoErr_TL}
\ig{\numAA}{pdf}{\figdirCheb fig_U4_err_cmpTL}
\caption{(Color online)
\tbf{Top}: 
Input spectral function and reconstructed spectral
functions using linear prediction for 
$N=200$ computed Chebyshev moments. 
We compare our proposal with 
the original proposal,\cp{ganahl14} 
where for the $b=0$ 
setup the expansion of the \tit{full} spectral 
function was extrapolated.
\tbf{Left}:
In the first case, we use the Chebyshev expansion
of $\wt A^>(\w)$ in the $b=-0.995$ setup (red dashed lines), 
and in the second, 
we use the Chebyshev expansion 
of the \tit{full} spectral function $A(\w)$ in the $b=0$ setup (solid green
line).
\tbf{Right}:
Error of these functions 
$|A\tl{reconst}(\w)-A\tl{input}(\w)|$ 
for both setups.
\tbf{Bottom}: The error  
$\max_{\w\geq0} |A\tl{reconst}(\w)-A\tl{input}(\w)|$
versus number $N$ of computed Chebyshev moments.
Here, we also show results for using the Chebyshev expansion
of $\wt A^>(\w)$ in the $b=0$ setup (light green crossed line). 
This is different from using the \tit{full} spectral 
function $A(\w)$ in the $b=0$ setup. 
Lower error levels only occur for much higher higher expansion
orders than shown in the panel.
}
\label{fig_SIAMcompareSetups}
\end{figure}

\figref{fig_SIAMcompareSetups} 
compares the approach of 
reconstructing the \tit{full} spectral funciton $A(\w)$ from $A^>(\w)$,  
using linear prediction for the expansion of $A(\w)$ in the $b=0$ setup, 
with the new approach of 
using linear prediction of $\wt A(\w)$ 
in the $b=-0.995$ setup.
We take the function of \figref{figSpecExample}
as input function that shall be reconstructed.
In the top panels of \figref{fig_SIAMcompareSetups}, 
we compare both setups
for $N=200$ computed moments 
that are then extrapolated to $N\gg 1000$ until
they converge to a value of $10^{-6}$.
We choose this comparatively small number of computed moments, as 
in MPS algorithms the number of moments that can be computed 
in a controlled way is strongly limited.\cp{wolf14} 

The upper left panel of \figref{fig_SIAMcompareSetups}
shows that already for $N=200$, our approach (dashed red line)
allows a very good reconstruction of 
the input function. In the upper right panel
we show the error of this reconstruction, which becomes maximal 
at the second peak of the input function and is of
order $10^{-2}$, i.e., a relative error of a few per cent.
The situation is very different for the extrapolation
scheme of the full $A(\w)$ that uses the $b=0$ 
setup. For $N=200$ computed
moments, large errors are observed in both top panels
of \figref{fig_SIAMcompareSetups}.

In the bottom panel of \figref{fig_SIAMcompareSetups},
we plot the maximal error, defined as 
$\max_{\w\geq0} |A\tl{reconst}(\w)-A\tl{input}(\w)|$,
versus different values of the number of computed moments $N$. 
An orders of magnitude reduction of the error is seen
upon using our over the previous approach. 
If one compares the expansion order $N$ for which
an error of $5\cdot 10^{-3}$ is reached 
($N\sim250$ in the $b\simeq-1$ setup, and $N=1200$ in the $b=0$ setup),
one recovers the factor $\sim 6$ that has been derived
in the previous section.

Only at very high expansion orders, that in practice can often not 
be reached by MPS computations, the original approach 
allows to reach smaller error levels for the presently studied generic example.
More examples are studied in \appref{secChebExamples}.

\section{Spectral functions for finite systems}
\label{secFS}

Let us now study the case of finite systems,
where a discretized representation of the spectral
function is used for reconstruction. 
The general previous arguments are still valid, 
but several technical details have to be taken into account. 
In particular, we suggest a new discretization scheme 
suited for reconstruction with Chebyshev expansions.
Such a discretization scheme can be used for problems
that allow to manipulate the discretization of the spectral function. 
This is \eg the case for impurity models, for which the discretization
of the input bath spectral function determines the discretization of the
spectral function. Still, the following discussion is also relevant 
to \eg finite lattice models for which the discretization 
is physically constrained.

To construct a discrete representation of a continuous function $A(\w)$,  
we employ the scheme that is used to discretize
the \tit{hybridization function} of impurity models in the 
numerical renormalization group.\cp{bulla08}
This proceeds as follows. 
For $L$ given discretization intervals 
$[\w_l, \w_{l+1}],\,\,l=1,...,L$, 
we compute discrete weights $V_l^2$ and eigenvalue positions $\e_l$ by
\eq{
V_l^2 & = \int_{\w_l}^{\w_{l+1}} d\w\,  A(\w), \quad   \label{Vlel}\\
\e_l  & = \frac{1}{V_l^2} \int_{\w_l}^{\w_{l+1}} d\w\, \w A(\w). \non
}
The first line associates a weight and 
the second line a representative energy
with an interval of energies $[\w_l, \w_{l+1}]$. 
For the energy, one could 
\eg take\cp{lin13} the simple average $\frac{1}{2}(\w_l+\w_{l+1})$.
\Eqref{Vlel}, by contrast, produces an average using the weighting function  
$\frac{1}{V_l^2}  A(\w)$, which attributes more weight
to peaks of $A(\w)$. 

We choose the left boundary of the first interval $\w_1$ and the right 
boundary of the last interval $\w_{L+1}$ such that
the distance $\w_{L+1}-\w_1$ is minimized but\cp{wolf14} 
\eq{ \label{wBounds}
\int_{\w_1}^{\w_{L+1}} d\w\,  A(\w) \geq 0.999 \int_{-\infty}^{\infty} d\w\,  A(\w),
}
where the integrand is non-negative, which guarantees that  
$[\w_1,\w_{L+1}]$ contains almost the complete support of $A^>(\w)$,
but minimizes \tit{finite-size} effects.
The intermediate values of the discretization intervals $\{ \w_2, ...,\w_L\}$ can be
chosen using a logarithmic discretization, as done in NRG.\cp{bulla08}
But if an \tit{unbiased} resolution is wanted,
one usually chooses a linear discretization\cp{ganahl14,wolf14}
\eq{
\w_l & = (l-1) \Delta\w + \w_1, \quad l = 2,...,L \non  \\
\Delta\w & = \tfrac{1}{L} (\w_{L+1}-\w_1) .  \label{linDiscr}
}

As Chebyshev polynomials do \tit{not} 
show an \tit{unbiased} energy resolution as
they oscillate much quicker at the boundaries 
of [-1,1] than in the center, 
the linear discretization will first resolve the 
finite size (discrete) structure 
close to the boundaries of $[-1,1]$.
We suggest to adapt the discretization to account for the
cosine mapping \eqref{intStretch} of the energy scale 
that is responsible for this phenomenon. 

Let us study the case of even $L$ (for odd $L$, see \appref{secDiscr})
and assume without loss of generality 
that we want as many intervals $\{\w_l,\w_{l+1}\}$
on the positive half-axis as on the negative half-axis,
which implies
\eq{ \label{wL2}
\w_{L/2} = 0.
}
As we already know $\w_{L+1}$ from \eqref{wBounds}, 
we only have to fix the intermediate
interval boundaries $\{\w_{L/2+1},...,\w_{L}\}$. 
We define
\eq{
\w_{L/2+l} & = a\big(\cos(\theta_{L/2} + l \Delta\theta)-b\big), \quad l = 1,...,L/2 \non \\
\Delta\theta &= \tfrac{2}{L} (\theta_{L+1}-\theta_{L/2}), \non\\
\theta_{L/2} & = \arccos b , \non\\
\theta_{L+1} & = \arccos(b + \tfrac{\w_{L+1}}{a}).    \label{cosDiscr}
}

Using these definitions, a discrete representation $A\tl{discr}(\w)$ of 
$A(\w)$ (in the sense that $A\tl{discr}(\w)\ra A(\w)$ for $L\ra\infty$) 
is given by 
\eq{
A\tl{discr}(\w) & = \bra{\psi_0} \delta(\w-H) \ket{\psi_0}, \non\\
H_{ll'} & = \e_l \d_{ll'}, \quad  l,l'=1,...,L \non\\
\ket{\psi_0}_l           & = V_l,     \label{specFuncSingle}
}
where $H\in \mathbb{R}^{L\times L}$ and $\ket{\psi_0}\in\mathbb{R}^L$,
and the parameters $\e_l$ and $V_l$ are given in \eqref{Vlel}.
This is consistent with definition \eqref{specFunc} if we realize
that this is a single-particle Hamiltonian for a particle that is 
in either of $\e_l$ energy states with probability $V_l^2$. 
The reference energy would be the ground state energy
of the vacuum $E\tl{ref}=0$. To obtain the step function behavior of $A^>(\w)$, 
we project out the positive energy contributions from the initial state $\ket{\psi_0}$.

\begin{figure}
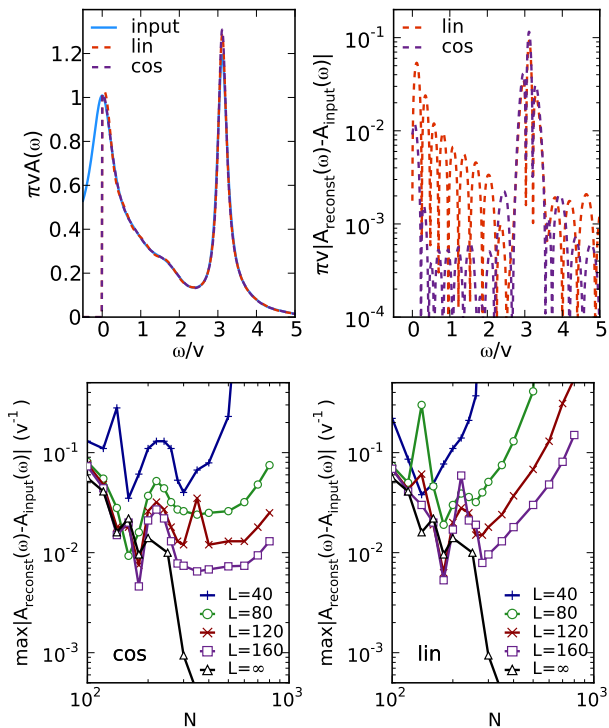

\ig{\numB}{pdf}{\figdirCheb fig_U4_rho_FS}
\ig{\numB}{pdf}{\figdirCheb fig_U4_rhoErr_FS}
\ig{\numB}{pdf}{\figdirCheb fig_U4_errFScos}
\ig{\numB}{pdf}{\figdirCheb fig_U4_errFSlin}
\caption{(Color online) 
Reconstruction of the spectral function 
of \figref{figSpecExample} represented
by the discrete Hamiltonian \eqref{specFuncSingle}.
\tbf{Top left}:
Input function and reconstructed functions using 
$L=80$, $N=200$, $a=100v$, $b=-0.995$.
We compare the linear discretization \eqref{linDiscr} 
with the cosine \eqref{cosDiscr}
discretization.
\tbf{Top right}: Difference of input and 
reconstructed functions of the top left panel.
\tbf{Bottom left}: 
The error  
$\max_{\w\geq0} |A\tl{reconst}(\w)-A\tl{input}(\w)|$
versus number $N$ of computed Chebyshev moments
using a cosine discretization.
\tbf{Bottom right}:
Error for linear discretization.
}
\label{figU4fs}
\end{figure}

In \figref{figU4fs}, we show the reconstruction
of spectral functions based on the linear prediction of 
the moments computed for $\wt A\tl{discr}^>(\w)$
using the operator-valued Chebyshev expansion 
presented in \secref{secChe} for the ``Hamiltonian" defined
in \eqref{specFuncSingle}.
This is analogous to the top left panel 
of \figref{fig_SIAMcompareSetups}, which treated the 
thermodynamic limit. 

For the finite-size system, the specific choice of discretization
is important and we compare the linear and the cosine
discretization in the top panels of \figref{figU4fs} for 
the expansion order $N=200$ and a system size $L=80$. 
From the large error at $\w=0$ for the linear
discretization (red dashed line) seen in the right top panel of \figref{figU4fs},
which was not present in the thermodynamic limit 
(red dashed line in right top panel of  \figref{fig_SIAMcompareSetups}), 
we conclude that the linear discretization 
starts resolving finite-size features close to $\w=0$ 
already for $N=200$. 
The lower panels then show how the error behaves
as a function of the number of computed moments for different
system sizes $L$. While the cosine discretization follows
the error of the thermodynamic limit quite closely for low
values of $N$ and lattice sizes of $L\geq80$, 
it almost saturates in a plateau for higher expansion orders, and only
starts increasing slightly for very high expansion orders. 
For the linear discretization, neither the close correspondence 
with the thermodynamic limit is observed, nor does the error only moderately
depend on the expansion order: 
Instead, the error increases exponentially for high values of $N$,
as then, finite-size features are inhomogeneously resolved. Both
features make it difficult to determine the value of $N$ for which the computation of 
Chebyshev moments should be stopped in order to obtain
a minimal error.

\section{Implications for MPS representations}
\label{secMPS}

What is the relevance of the previous results for
matrix product state (MPS) based 
computations of the spectral function 
for a given matrix product operator (MPO) $H$?\cp{schollwock11} 
Repeated MPO operations on MPS create entanglement, 
which eventually makes manipulating and storing  
MPS computationally very costly. 
Manipulations, such as applying 
$H$ to states $\ket{t_n}$ in the recursion \eqref{rec}, 
or performing subsequent time evolution 
steps $e^{-iH\Delta t}$, can therefore
only be carried out up to a certain 
recursion order $n$ or time $t$, before
hitting an \tit{exponential wall} in computation cost.
For time evolution algorithms, this has long been known,\cp{gobert05,eisert06} 
but this also limits computations using 
the Chebyshev recursion.\cp{wolf14}

In the following, we show that the method 
introduced in the previous sections 
outperforms the previous approach:\cp{ganahl14} 
it extracts more spectral information from $H$ 
when creating the same amount of entanglement or,
which is equivalent up to technical details of the algorithm, using the same 
computation time.

As an example, we compute the spectral function of the single
impurity Anderson model (SIAM), 
which serves as a common benchmark\cp{raas04,ganahl14,ganahl14i,wolf14,wolf14i}
and is highly relevant as it is at the core of dynamical mean-field theory (DMFT).\cp{metzner89,georges92,georges96,kotliar06}

The Hamiltonian of the SIAM is given as, 
\eq{
H\th{SIAM} & = H\tl{imp} + H\tl{bath} + H\tl{hyb},  \label{Hsiam} \\
H\tl{imp} & = U \left(n_{0\uparrow}-\tfrac12\right)\left(n_{0\downarrow}-\tfrac12\right), \non\\
H\tl{bath} & = \sum_{l=1}^{L_b}\sum_{\sigma} \epsilon_{l} c\dag_{l\sigma}c_{l\sigma},\non\\
H\tl{hyb} & = \sum_{l=1}^{L_b}\sum_{\sigma}\left(V_{l} c\dag_{0\sigma}c_{l\sigma}+\text{H.c.}\right). \non
} 
By a unitary transform effected by Lanczos tridiagonalization,
this can be mapped on the so-called \tit{chain} geometry.  
But as this leads to higher entanglement, we simply order bath states
by their potential energy which directly gives a one-dimensional array
that can be treated with MPS.\cp{wolf14i}
We solve the model for the semi-elliptic bath density of states
\eq{  \label{eqSemiEll}
-\frac{1}{\pi} \tx{Im} \Lambda(\w) = \frac{1}{2v\pi} \sqrt{4 - \Big(\frac{\w}{v}\Big)^2},
}
which is discretized according to the procedure discussed in \secref{secFS},
and then yields the parameters $\e_l$ and $V_l$. It is important to realize that here,
we discretize the \tit{bath hybridization} function whereas in \secref{secFS},
we discretized the \tit{spectral function}. While \secref{secFS} did this
to illustrate the effect of discretization for a toy model for which the spectral function
was known from the beginning, in the present case,  a true many-body compuation is 
involved. In the present case, the relevant discretization
parameter is the bath size $L_b=L-1$, and no longer the system size.\cp{wolf14} 

We compute the spectral function \eqref{eqFull} of 
the impurity Green's function, where the initial states
are single-particle excitations of the ground state:  
$\ket{\psi_{0\s}^>} = c_{0\s}\dag \ket{E_0}$ and $\ket{\psi_{0\s}^<} = c_{0\s} \ket{E_0}$. 
As we consider the particle-hole and spin-symmetric case of \eqref{Hsiam}, we only
need to compute one Chebyshev recursion; 
to be precise: $\ket{\psi_0} = c_{0\uparrow}\dag \ket{E_0}$.
We compare our results with dynamic DMRG results from \oc{raas04}, which 
are believed to be highly reliable. In particular, we compare computations
in the formerly suggested setup\cp{holzner11,ganahl14} that 
uses the Chebyshev recursion for $b=0$ in 
\eqref{eqShiftScale} and reconstructs the \tit{full} 
spectral function $A(\w)$ using linear prediction,\cp{ganahl14} 
and the one suggested here, that uses $b=-0.995$ 
and reconstructs the \tit{shifted} spectral function $\wt A(\w)$ using linear prediction.
\begin{figure}
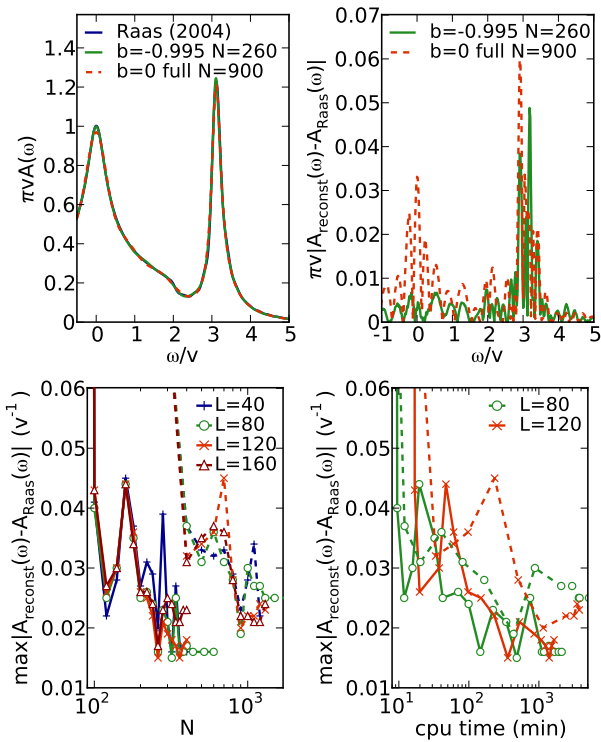

\ig{\numB}{pdf}{\figdirCheb fig_MPS_SIAMU4L80}
\ig{\numB}{pdf}{\figdirCheb fig_MPS_SIAMU4L80_err}
\ig{\numB}{pdf}{\figdirCheb fig_MPS_SIAMU4_err}
\ig{\numB}{pdf}{\figdirCheb fig_MPS_SIAMU4_errVScput}
\caption{(Color online) 
Comparison of MPS computed spectral functions
in the two setups studied in the previous sections,
with data by \ct{raas04}. Solid lines refer to the new method that uses
$\wt A(\w)$, dashed lines refer to 
the method that uses the \tit{full} spectral function $A(\w)$. 
\tbf{Top left}: For $L=80$ 
and two exemplary expansion orders.
\tbf{Top right}: Errors of comparison in top left panel.
\tbf{Bottom left}: Plot of the maximum error versus
expansion order $N$. 
\tbf{Bottom right}: Plot of the maximum error versus computer time.
}
\label{figMPSU4}
\end{figure}

In the top left panel of \figref{figMPSU4}, 
we show computations of the spectral function of the SIAM for 
$L=80$ for $N=260$ in the $b=-0.995$ setup, and $N=900$ in the $b=0$ setup 
and compare it with the result of \oc{raas04}.
We choose these two expansion orders, as they lead to a comparable
maximum error, as shown in the top right panel of \figref{figMPSU4}.
In the $b=-0.995$ setup, this maximum error is slightly smaller.
Around $\w=0$, by contrast, the error in the $b=-0.995$ 
setup is much smaller. If we compare the computation time that is
needed to reach this precision ($\tx{max}|A\tl{reconst}(\w)-A\tl{input}| \simeq 0.015/v$), 
we find that the $b=-0.995$ setup
required $\sim$\SI{145}{min} whereas the $b=0$ setup 
required $\sim$\SI{434}{min}. If one makes this comparison
for a slightly larger error ($\tx{max}|A\tl{reconst}(\w)-A\tl{input}| \simeq 0.025/v$), 
realized for expansion order $N=120$
for the $b=-0.995$ setup, and for expansion order $N=200$ for the 
$b=0$ setup, the comparison in computation times reads
$\sim$\SI{12}{min} versus $\sim$\SI{160}{min}. 

When studying the convergence of the maximum error 
with respect to expansion order $N$
in the lower left panel of \figref{figMPSU4},
we see that this is, after a sharp decrease for low expansion orders, not 
monotonously decreasing. The previously mentioned choices,  
$N=260$ in the $b=-0.995$ setup and $N=900$ in the $b=0$ setup, 
both correspond to a minimum in the oscillations, as seen
when inspecting the green solid (dashed) lines for the 
$b=-0.995$ ($b=0$) setup.
The non-monotonicity makes general comparisons for the 
speedup difficult. But the lower right panel of \figref{figMPSU4} 
still shows that with only a few exceptions,
the solid ($b=-0.995$) lines are always clearly below the dashed ($b=0$)
lines. The logarithmic abscissa therefore indicates a high speedup.

\section{Comparison to time evolution}
\label{secTimeEvol}

It is interesting to compare the efficiency
of the available MPS algorithms to extract spectral information 
from $H$. Candidates are, aside from the dynamic DMRG,\cp{jeckelmann02}
which is believed to be computationally highly costly, 
time evolution and 
recursive algorithms. The latter are, in particular, 
expansions in Chebyshev polynomials\cp{holzner11}
and the Lanczos algorithm.\cp{garcia04,dargel12}
Lanczos is numerical instable as the basis that it spans
looses its orthogonality for high numbers of iterations.\cp{arbenz12}
This seems to to disqualify Lanczos as a high-performing candidate.
Therefore the main question is whether the Chebyshev
recursion can more efficiently extract spectral
information from $H$ than time evolution algorithms. 

To answer this question, in the following, we exploit the fact that for $b=0$
in the limit $a\rightarrow\infty$, the Chebyshev expansion
becomes a Fourier expansion, and the Chebyshev 
states directly describe the time evolved system. 
This is different from the procedure of computing
the time dependence of a Green's function
via a Fourier transform of its spectral function.\cp{talezer84,leforestier91,holzner11,halimeh15}
For comparison, we summarize the latter 
technique in \appref{secStanTimeEvol}.

The approximate equivalence
of the Chebyshev recursion to time evolution
can be used as a novel time evolution algorithm.
This is interesting as for long-range interacting  
Hamiltonians $H$, the MPO representation
of $e^{-iHt}$ is not available, or only approximately.\cp{zaletel14} 
Although it is possible to use so-called Krylov 
algorithms for such problems,
this requires some programming effort, and 
is in general believed to be numerically rather inefficient
as compared ot other time evolution algorithms.  
Long-range interacting problems appear \eg if mapping
a two-dimensional system on a one-dimensional chain, 
or in the solution of a SIAM using a \tit{star} geometry\cp{wolf14i} as in \eqref{Hsiam}.

\subsection{Statement of approximate equivalence}

The time evolution of a state $\ket{\psi_0}$
\eq{  \label{eqSchEQ}
\ket{\psi(t)} = \exp(-iHt) \ket{\psi_0},
}
can be approximately linked to the sequence of Chebyshev vectors 
generated by starting from $\ket{\psi_0}$ as follows.

Choose a reference energy $E\tl{ref}$ 
in \eqref{eqShiftScale} that is characteristic for the initial state
of the time evolution and the Chebyshev recursion. When computing
the time evolution of the Green's function with $\ket{\psi_0} = c\dag\ket{E_0}$,
one chooses $E\tl{ref}=E_0$, if $\ket{\psi_0}$ is not an eigenstate, 
we choose $E\tl{ref} = \bra{\psi_0} H \ket{\psi_0}$.  

Then define   
$\ket{\phi(t)} =  \exp(iE\tl{ref}t) \ket{\psi(t)}$
and $\mcal{H} = (H-E\tl{ref})/a$ as in \eqref{eqShiftScale} in the $b=0$ setup.
Here $a$ has the meaning of an inverse time step of unit energy. 
With these definitions, \eqref{eqSchEQ} reads as
\eq{ \label{eqEuler}
\ket{\phi(t)} 
& = \exp(-ia\mcal{H}t) \ket{\psi_0}  \non\\
& = \big(\cos(a\mcal{H} t) - i \sin(a\mcal{H} t) \big) \ket{\psi_0}  \non\\
& \equiv \ket{\phi\tl{cos}(t)}  -  i \ket{\phi\tl{sin}(t)}. 
}
Let us discretize time by defining $t_n = \frac{n}{a}$, then 
\begin{subequations}
\label{eqEulerDiscr}
\eq{
  \ket{\phi\tl{cos}(t_n)}  =  \cos(n \mcal{H}) \ket{\psi_0},  \\
  \ket{\phi\tl{sin}(t_n)}   =  \sin(n \mcal{H}) \ket{\psi_0}.   
}
\end{subequations}

We now want to compute the action of $\cos( n \mcal{H}) \ket{\psi_0}$
on $\ket{\psi_0}$ using a recursion that only involves the action of $\mcal{H}$.
This is not possible with the standard recursion for the cosine function, 
as shown in \appref{seqStandardRec}.

Let us instead consider the action of 
the Chebyshev polynomials  
\eq{
 T_n(\mcal{H}) =  \cos(n\arccos(\mcal{H})) 
}
on $\ket{\psi_0}$.  This action
approximately reproduces the action 
of the plane cosine function, if we consider every 4th iteration, \ie introduce
the new index $n'=4n$, $n\in\mathbb{N}$:
\eq{  
T_{n'} (\mcal{H}) \ket{\psi_0} 
& = \cos\Big(n'\Big(\frac{\pi}{2} - \mcal{H}\Big)\Big) \ket{\psi_0} + \e(n') \ket{\psi_0} \label{eqApprox} \\
& = \cos\Big(n' \mcal{H}\Big) \ket{\psi_0} + \e(n') \ket{\psi_0}. ~ n' = 0,4,8,...  \non
}
In the first line, we used the Taylor 
expansion $\arccos(\mcal{H}) = \frac{\pi}{2} - \mcal{H} + \frac{1}{6} \mcal{H}^3 + \cdots$,
that leads to the error function $\e(n')$ (\appref{secError}),
and in the second line, 
we used $n' \frac{\pi}{2}=2\pi n$, $n\in\mathbb{N}$, which obviously drops out of the argument of the cosine
(also see \appref{secShiftedCos}).

The error $\e(n')$ is bounded by \eqref{errorTimeApp}
 \eq{
|\e(t_{n'})|
& = \frac{t}{t\tl{err}} \qquad \tx{if} \quad  t < t\tl{err}  \non \\
t\tl{err} & = \frac{a^2}{\sigma^3}      \label{errorTime}
}
where $\sigma$ is the spectral width of the 
initial state $\ket{\psi_0}\equiv\ket{\psi_0}$ around $E\tl{ref}$,
\eq{ \label{spectralWidth}
\sigma = \max_{\ket{E_k} \in \ket{\psi_0}} |E_k-E\tl{ref}|,
}
where ``$\ket{E_k} \in \ket{\psi_0}$" refers to the 
decomposition of the initial state in 
eigenstates $\ket{E_k}$ of $\mathcal{H}$
\eq{
\ket{\psi_0} = \sum _k c_{k}  \ket{E_k}.
}
The ``spectral width" $\sigma$ is usually small 
compared to reasonably high values of the inverse time step $a$. 
If one is unsure of whether $a$ was chosen large enough, one 
reruns a calculation with a higher value of $a$ and checks convergence.

We can now compute the time evolution 
\eq{
\ket{\phi\tl{cos}(t_{n'})} & = \ket{\phi\tl{cheb}(t_{n'})} + \e(t_{n'}) \ket{\psi_0}, \non\\
\ket{\phi\tl{cheb}(t_{n'})} & = T_{n'}(\mcal{H}) \ket{\psi_0}, \quad n'=0,4,8,\dots     \label{cheTimeEvol}
}
via the recursion \eqref{rec}
\eq{
\ket{\phi\tl{cheb}(t_n)} & = 2\mcal{H} \ket{\phi\tl{cheb}(t_{n-1})}  - \ket{\phi\tl{cheb}(t_{n-2})},  \non \\
\ket{\phi\tl{cheb}(t_1)} & = \mcal{H} \ket{\psi_0}, \quad n=0,1,2,\dots. \label{recTime}
}

\begin{figure}[!h]
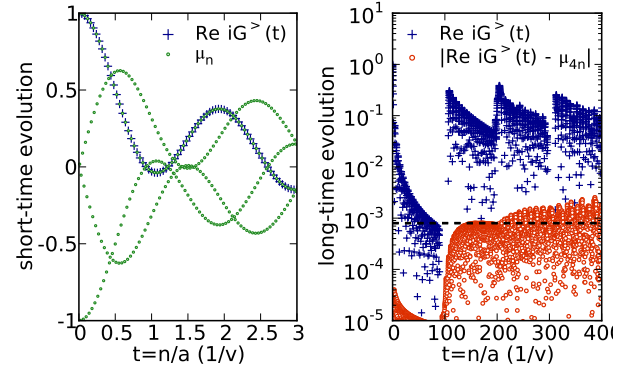

\ig{\numB}{pdf}{\figdirCheb fig_evolU4}
\ig{\numB}{pdf}{\figdirCheb fig_evolU4err}
\caption{(Color online) 
\tbf{Left}:
Time evolution of a particle created on the first
site of a chain of length $L=100$ with hopping $v=1$ that is 
$\ket{\psi_0} = c_0\dag\ket{\tx{vac}}$. 
We compare the time evolution of the  Green's function 
$iG^>(t) = e^{iE_0t}\ip{\psi_0}{\psi(t)}$ (shown as blue crosses) 
with the Chebyshev moments  
$\mu_n = \ip{\psi_0}{\psi_n}$ (shown as green dots) 
obtained when computing the 
recursion $\ket{\psi_n} = 2H/a \ket{\psi_{n-1}} - \ket{\psi_n-2}$.
Hopping amplitudes $v_l$ are obtained from the discretization 
of the spectral function of \figref{figSpecExample}.
A qualitatively equivalent behavior is obtained
for a chain with homogeneous hopping $v_l=v$.
\tbf{Right}: Long-time evolution of the Green's 
function (blue crosses), and difference of the 
Green's function and the Chebyshev moments (red dots).
The horizontal dashed line marks the the prefactor 
of the error estimate \eqref{errorTime}, which is computed
as $\sigma^3/a^2=8\cdot10^{-4}$.
}
\label{figEvol}
\end{figure}
\subsection{Numerical examples}
\subsubsection{Single-particle computation for SIAM}

\figref{figEvol} shows the numerically exact time evolution of a single 
particle created on the first site $\ket{\psi_0} = c_0\dag\ket{E_0}$ 
of a chain of 100 lattice sites. The spectral width therefore is $\sigma=2v$. 
The left panel of  \figref{figEvol} plots the Chebyshev moments
 $\mu_n = \ip{\psi_0}{t_n}$ obtained with $a=100v$ and the time evolution 
of the corresponding Green's function 
$iG^>(t) = e^{iE_0t}\ip{\psi_0}{\psi(t)}$ 
for the time step $\frac{4}{a}$.
Every forth Chebyshev moment agrees with a value of the 
Green's function. In the right panel, 
we show the long time behavior of $G^>(t)$
and the difference $G^>(4n/a) - \mu_{4n}$.
The difference is seen to be clearly below
the conservative upper bound \eqref{errorTime}, it remains of the order 
of $\sigma^3/a^2=8\cdot10^{-4}$ up to very high times that correspond to 100 hopping processes ($t=100/v$).

\subsubsection{MPS computation for SIAM}

\begin{figure}[!h]
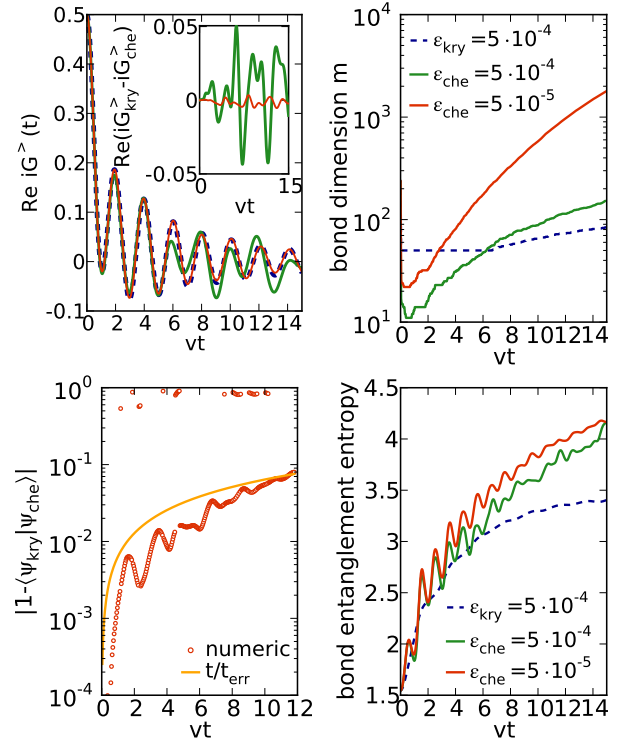

\ig{\numB}{pdf}{\figdirCheb fig_kry_SIAMU4_GgtrL40}
\ig{\numB}{pdf}{\figdirCheb fig_kry_SIAMU4_mL40}
\ig{\numB}{pdf}{\figdirCheb fig_kry_SIAMU4_overlapL40}
\ig{\numB}{pdf}{\figdirCheb fig_kry_SIAMU4_entL40}
\caption{(Color online) 
Time evolution of the single particle excitation
$c_{0\uparrow}\dag\ket{E_0}$ in the half-filled single impurity 
Anderson Model \eqref{Hsiam} with semi-elliptic density
of states of half-bandwidth $2v$ and interaction $U/v=4$
for $L=40$. 
Computations using an MPS Krylov algorithm
with error tolerance $\varepsilon\tl{kry}=5\cdot10^{-4}$ and
the Chebyshev recursion \eqref{cheTimeEvol}
for different error tolerances 
$\varepsilon\tl{che}=5\cdot10^{-4}$ 
and $\varepsilon\tl{che}=5\cdot10^{-5}$.
\tbf{Top left}: Time evolution of Greens's function. 
Both algorithms produce the same result upon using the
smaller error tolerance for the Chebyshev algorithm.
The legend is found in the top right panel.
\tbf{Top right}: Maximal bond dimension, located at the central bond. 
The Krylov time evolution leads to a much smaller  
maximal bond dimension, as its computation produces 
a faithful result already with the relatively high error tolerance of 
$\varepsilon\tl{kry}=5\cdot10^{-4}$, for which the Chebyshev 
algorithm shows strong errors in the Green's function.  
\tbf{Bottom left}:
Difference of overlap of Chebyshev and Krylov
evolved states, comparing the $\varepsilon\tl{che}=5\cdot10^{-5}$
with the $\varepsilon\tl{kry}=5\cdot10^{-4}$
computation.
The difference of overlap is bounded by 
the analytical prediction of \eqref{errorTime}, except for few 
exceptions that lie above it. These exceptions are of purely numerical
origin as they are not visible in any other quantity. For the highest times shown, 
truncation errors have accumulated so much that the analytical prediction 
starts to fail.
\tbf{Bottom right}:
Bond entanglement entropy at center bond.  
}
\label{figEvolSiamMPS}
\end{figure}

We now study the time evolution
of the SIAM \eqref{figEvolSiamMPS} in the star geometry\cp{wolf14i} for 
the single-particle excitation $\ket{\psi_0} = c_{0\uparrow}\dag\ket{E_0}$
of the half-filled ground state $\ket{E_0}$. 
The left panel shows the time evolution of the corresponding
Green's function, computed with an MPS Krylov algorithm
that imposes the error bound
\eq{
\big\vert\big\vert\,\ket{\psi(t+\D t)} - \exp(-i H \D t) \ket{\psi(t)} \,\big\vert\big\vert < \ve\tl{kry}, \non
}
for a time step of $\Delta t = \frac{4}{a}$.
An error bound of $\ve\tl{kry}=5\cdot10^{-4}$ suffices to reliably 
compute times up to $15/v$. 

In the MPS implementation of the Chebyshev recursion,
we fix the global truncation error per iteration step, as
discussed in \oc{wolf14} 
\eq{ \label{truncChe}
\big\vert\big\vert\, \ket{t_n} -  (2\mcal{H} \ket{t_{n-1}} - \ket{t_{n-2}}) \,\big\vert\big\vert < \ve\tl{che}.
}
To achieve this, two options are available. If during the variational 
compression of $(2 \mcal{H} \ket{t_{n-1}}-\ket{t_{n-2}})$, 
the truncation error exceeds $\ve\tl{che}$,  
even when choosing a better and better guess state, one can either
directly increase the bond dimension, or reduce the truncated
weight \tit{per bond}, which indirectly increases the bond dimension. 
While for the setup in \oc{wolf14}, there were reasons to 
choose the former option, here we choose the latter as 
our Krylov algorithm uses a similar adaption.

We compare the results of the Krylov algorithm 
with the Chebyshev algorithm \eqref{cheTimeEvol}. 
In the top left panel of \figref{figEvolSiamMPS}, we plot the Green's function
$iG^>(t) = e^{iE_0t}\ip{\psi_0}{\psi(t)}$.  
If imposing the same error tolerance $\ve\tl{che}=\ve\tl{kry}$, we  
obtain agreement of both algorithms only for short times. 
Only a much smaller tolerance for the Chebyshev algorithm
$\ve\tl{che}=\frac{1}{10}\ve\tl{kry}$ leads to agreement also for long times. 
We conclude that error accumulation in the Chebyshev recursion
is much worse conditioned than in the time evolution algorithm,  and
even worse than what could be expected from the four ``auxiliary
steps" made in \eqref{recTime} between each ``physical time step":
imposing a tolerance $\ve\tl{che}=\frac{1}{4}\ve\tl{kry}$ for the Chebyshev
recursion is not sufficient to produce comparable results. 

The reduced error tolerance  $\ve\tl{che}=\frac{1}{10}\ve\tl{kry}$ for the Chebyshev recursion  
comes at the price of an order of magnitude increase 
in the bond dimension compared to the Krylov algorithm,  
as shown in the top right panel of \figref{figEvolSiamMPS}. 
But also for $\ve\tl{che}=\ve\tl{kry}$, the Chebyshev recursion
needs higher bond dimensions than the Krylov algorithm.
The lower left panel compares 
the overlap of the Chebyshev-evolved and the Krylov-evolved
states by plotting $|1 - \ip{\psi\tl{cheb}}{\psi\tl{kry}} /\ip{\psi\tl{cheb}}{\psi\tl{cheb}}|$.
With only few exceptions, this quantity is bounded by the
theoretical prediction of \eqref{errorTime}, when setting $\sigma=4v=U$.
The exceptions are artifacts of the detailed implementation of 
the algorithms as the key observable $G^>(t)$ is correctly computed, 
but still their existence suggest that the implementation can be improved.
Ignoring these exceptions, we see that the normalized overlap $\ip{\psi\tl{cheb}}{\psi\tl{kry}}$
deviates from one only by a few percent even for long times. But these few percent
come with a considerable growth of the entanglement entropy, as can be concluded
by inspecting the lower right panel of \figref{figEvolSiamMPS}. There,
already the $\ve\tl{che}=\ve\tl{kry}$ case shows a considerably increased entropy.

Aside from the two preceding fundamental reasons (different error accumulation,
small difference of states), the increased bond dimensions
in the Chebyshev algorithm can also be related to 
a purely technical question: the variational 
compression\cp{schollwock11} in each Chebyshev iteration 
produces a state that fulfills \eqref{truncChe},  
but might be a state with unnecessarily high 
bond dimension $m$. 
Similarly to the DMRG ground state optimization algorithm,
also variational compression can get 
stuck in \tit{local minima}. Currently, we use White's \tit{mixing factor}\cp{white05}
to avoid this. A recent publication suggests an even better strategy  
and explains these problems concisely.\cp{hubig15} 

\begin{figure}[!h]
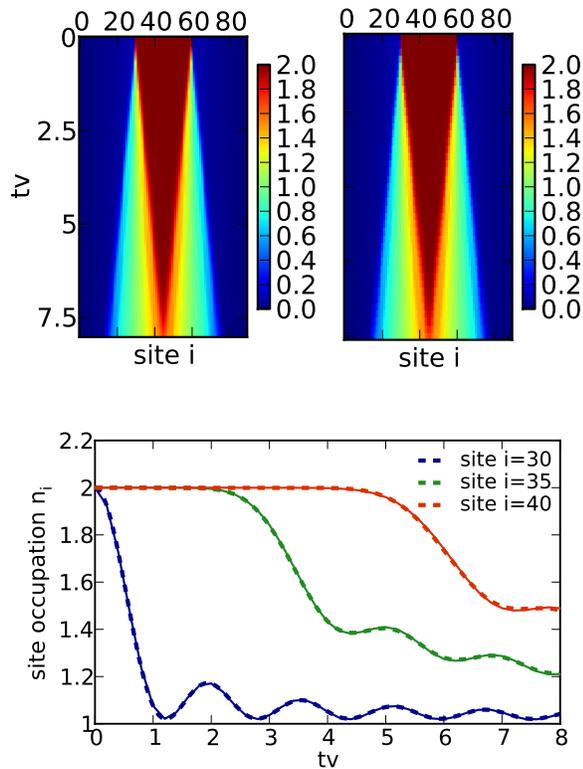

\ig{\numAA}{pdf}{\figdirCheb fig_MPS_hubbardDens}
\ig{\numAA}{pdf}{\figdirCheb fig_MPS_hubbardDensSingle}
\caption{(Color online) 
Time evolution  
of a one dimensional fermionic Hubbard model on $L=90$ sites, with an interaction of $U/v=4$
and nearest neighbor hopping $v$
starting from a product state (double occupation in the center of the system).
\tbf{Left}: Chebyshev computation using $\ve\tl{che}=0.0001$.
\tbf{Right}: Krylov computation using $\ve\tl{kry}=\ve\tl{che}$.
\tbf{Bottom}: Detailed comparison for the occupation
of specific sites. Chebyshev results are shown as dashed lines,
Krylov results are shown as solid lines. Deviations are smaller than 1\%.
}
\label{figEvolHub}
\end{figure}

In general the subspace of the Hilbert space that is needed to be 
faithfully described in order to measure the spectral function can be spanned
using different basis states. In principle, the most efficient
spanning would be provided by the Lanczos algorithm, as the latter 
provides \tit{orthogonal} states. But it is impractical due to numerical instability.
The basis states provided by time evolution or the Chebyshev recursion are not orthogonal
to each other, but can be stably generated. 
The numerical evidence discussed in the previous paragraphs 
indicates that the Chebyshev 
recursion generates a much higher
entangled basis of this subspace than a time evolution algorithm:
it extracts less spectral information when fixing a maximal entanglement entropy. 
But these arguments directly hold only for the ``$b=0$ setup" 
of the Chebyshev recursion, in which it is transparently comparable with a time evolution
algorithm as there is a one-to-one correspondence of time evolution steps and
iterations of the recursion. 

Sections \ref{secTL}--\ref{secMPS} of this paper 
showed that the $b=0$ setup is the computationally least favorable setup of
the Chebyshev recursion, and a $b\simeq-1$ setup much better. 
Still the gains in computation time of the $b\simeq-1$ setup over the $b=0$ setup 
shown in the bottom right panel of \figref{figMPSU4} seem not to be sufficient
to compensate the clear inferiority of the Chebyshev method shown in the upper
right panel of \figref{figEvolSiamMPS}. A definitive statement is difficult due to the non-monotonic 
behavior of the error in \figref{figMPSU4} and due to the fact that 
such a comparison is strongly affected by 
the details of the implementation of the algorithms, 
and not only by the principle nature of
how strongly entangled its resulting basis states are. 
For this reason, the discussion
on the most efficient method for computing spectral functions using MPS cannot be
generally considered settled.

\subsubsection{Expansion in Hubbard model}

Finally, we study the time evolution of the one-dimensional Hubbard model
\eq{
H\th{Hubbard} = U \sum_l \left(n_{l\uparrow}-\tfrac12\right)\left(n_{l\downarrow}-\tfrac12\right) \non\\
- v\sum_{l\s}( c\dag_{l\sigma}c_{l+1\sigma} + \tx{h.c.}),
} 
starting from a product state with doubly occupied sites in the center of the
system, and evolving this state at interaction $U/v=4$, as shown in \figref{figEvolHub}.
We obtain very good agreement of the Krylov and the 
Chebyshev algorithm, although there is no rigorous \tit{a priori} reason,
for which the initial product state should have a narrow spectral width, \ie small $\sigma$
in the sense of \eqref{spectralWidth}, as was the case for the single particle
excited initial state. On the other hand, for example
in the many studies on 
the \tit{eigenstate thermalization hypothesis}\cp{deutsch91,srednicki94} 
it is a frequently met assumption 
that for typical initial states the energy distribution around its 
mean value is extremely narrow, with a width of the
order of the single-particle energy scale (see \eg \oc{rigol07} Fig. 3b).

\section{Conclusion}
\label{secCon}

We started by linking \tit{linear prediction}
to \tit{analytic continuation}, which explains why 
linear prediction is a reasonable 
method to extrapolate both
Fourier and Chebyshev expansions of spectral functions. 
In order to apply linear prediction, 
we introduced a new method 
to avoid the algebraic convergence 
of Chebyshev moments 
(expansion coefficients) of generic step-like spectral functions. 
This amounts to a particular redefinition of the series expansion 
that is based on a subtraction of the Chebyshev moments of  
a self-consistently determined step function.

We then showed that this allows to reduce the expansion order 
by a factor $\frac{1}{6}$ as compared to the existing method.\cp{ganahl14} 
For linearly scaling algorithms, as in exact diagonalization,\cp{weisse06,lin13} this means 
a reduction of computation time of the same factor. 
But also for matrix product state computations high speedups are obtained. 
Furthermore, we showed how to adapt the discretization 
of hybridization functions of impurity models to the Chebyshev method.

Finally, we showed the approximate equivalence of the Chebyshev recursion to 
time evolution in a certain limit. This lead to a novel time evolution 
algorithm and allowed to transparently compare standard  
time evolution and the Chebyshev recursion
in how efficient they extract spectral information from an operator $H$.
For exact representations, the Chebyshev recursion is superior
to time evolution as the latter is equivalent to the least favorable setup of the Chebyshev
expansion, which can be improved by the previously mentioned factor $\frac{1}{6}$. 
For matrix product state representations, our results indicate that 
the Chebyshev expansion is inferior:  we observe 
a much higher entanglement production
in the Chebyshev recursion than in standard time evolution.   
We identify as main reason for this an unfavorable 
error accumulation in the Chebyshev
recursion that requires computations at higher accuracy. 
So while in the history of the solution of differential equations 
for non-time periodic problems, Chebyshev
expansions replaced Fourier expansions 
in the course of time,\cp{boyd01} 
in the matrix product state context such a transition now seems unlikely.
Still the Chebyshev recursion provides an easy-to-implement 
and straight-forward way to compute spectral functions.

Relevant applications of the results of this paper are the computation
of conductivities,\cp{garcia14} the computation 
of time evolution of long-range interacting systems,\cp{zaletel14}
and in particular, the challenging solution of dynamical mean-field theory.\cp{ganahl14,wolf14}
For example, the latter can usually not be accessed by combining analytical and numeric techniques 
as recently done for the Hubbard model in \oc{seabra14}.

\section{Acknowledgements}

FAW acknowledges fruitful discussions with
D.~Braak, M.~Eckstein and R.~Leike, 
and support by the research unit FOR 1807 of the DFG.

\section*{Appendix}
\appendix

\section{Convergence speed}
\label{secConv}

Analogously to \oc[Chap.~2.9]{boyd01}, we give the argument
for the speed of convergence of the Chebyshev sequence,
computed with the non-weighted inner product of \eqref{eqScalarProd}
\eq{
\mu_n 
& = \int_{-1}^1 dx f(x) T_n(x) \non\\
& = \int_0^\pi d\theta f(\cos\theta) \cos(n\theta)\sin\theta \non\\
& = \tx{Re} \int_0^\pi d\theta \wt f(\cos\theta) e^{i n\theta},
}
where $\wt f(\theta) = f(\cos\theta) \sin\theta$. We can then do 
$k$ partial integrations, if  $\wt f(\theta)$ is  $k$ times differentiable
\eq{\label{bound}
\mu_n 
& = \tx{Re} \Big[ - \sum_{j=1}^k  \Big( 
                     \Big(\frac{i}{n}\Big)^j e^{in\theta} \wt f^{(j-1)}(\theta) \Big\vert_0^\pi \Big) \non\\
& \qquad +  \Big(\frac{i}{n}\Big)^k \int_0^\pi d\theta \wt f^{(k)}(\theta) e^{in\theta} \Big],
}
where $\wt f^{(j)}(\theta)$ denotes the $j$th derivative of $\wt f(\theta)$.
If $\wt f^{(j)}(0)=\wt f^{(j)}(\pi)=0$ for $j=0,\dots, k-1$, which is fulfilled for typical
single-particle spectral functions as in \figref{figIntegrand}, 
and if $\wt f^{(k)}(\theta)$ is integrable, 
\eqref{bound} constitutes an upper bound $\mcal{O}\big(\frac{1}{n^k}\big)$ 
for the sequence $\mu_n$.

\section{Examples for linear prediction of Chebyshev expansions}
\label{secChebExamples}

In \secref{secTL}, we compared the reconstruction of 
a spectral functions using its extrapolated
(linearly predicted) Chebyshev expansion. 
We focussed on a typical example for this discussion,
given by the $U/v=4$ spectral function of the half-filled SIAM with
semi-elliptic bath density of states, which is shown in 
the top panel of \figref{figSpecExample}. 

In this appendix, we support the arguments of \secref{secTL}
by showing further generic examples. 
Starting from a step-like input function $A^>(\w)$, we again compare
the two reconstructions based on (i) 
linearly predicting the ``subtracted'' 
spectral function $\wt A^>(\w)$ of \eqref{eqSub} and (ii) linearly
predicting the ``full'' 
(summed particle and hole contributions) 
spectral function $A(\w)$ of \eqref{eqFull}.

\begin{figure}
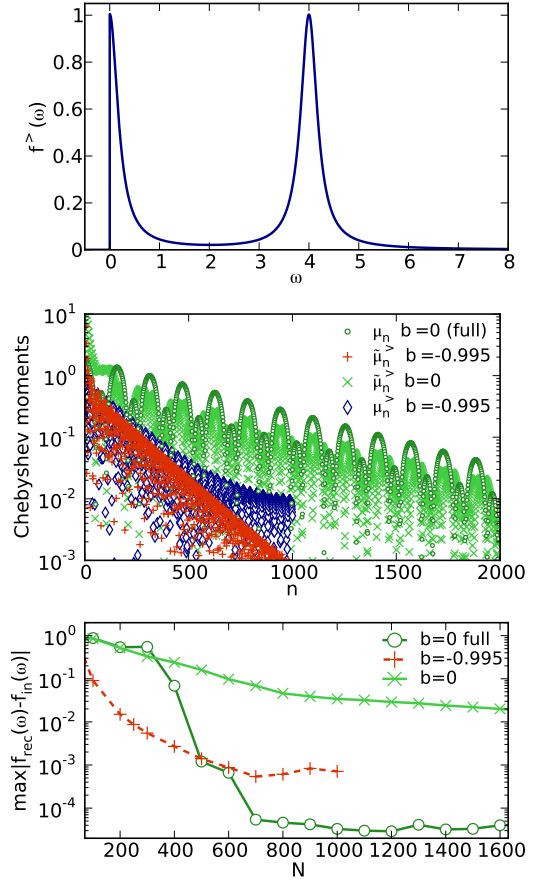

\ig{\numA}{pdf}{\figdirCheb fig_lor_rho}
\ig{\numA}{pdf}{\figdirCheb fig_lor_mu}
\ig{\numA}{pdf}{\figdirCheb fig_lor_err_cmpTL}
\caption{(Color online) 
\tbf{Top}: Test function consisting of two Lorentzians \eqref{testf}. 
\tbf{Center}: Corresponding Chebyshev moments
in the three different setups $\mu^>$, $\wt\mu^>$ and $\mu$,
analogously to the bottom panel of \figref{figIntegrand}.
\tbf{Bottom}: Error of reconstructed spectral function,
analogously to the bottom panel of \figref{fig_SIAMcompareSetups}.
All of this is for $a=100$.
}
\label{figLor}
\end{figure}

To consider generic cases, we study functions
that show ``features" at $\w=0$ and at some distance,
of order of the single-particle bandwidth, away from it. 
The most natural choice for constructing such functions
are superpositions of
(non-normalized) Lorentzians and Gaussians
\eq{ \label{testf}
f^{>}(\w) & = 
\left\{
\begin{array}{ll}
0 & \tx{for }  \w < 0 \\
\sum_{\w_0\in\{0,4\}} h(\w,\w_0) & \tx{else,}
\end{array}
\right.\\ 
h_l(\w,\w_0) & = \frac{\eta^2}{(\w-\w_0)^2+\eta^2},   \non\\
h_g(\w,\w_0) & = e^{-\frac{(\w-\w_0)^2}{2\eta^2}}.  \non
}
The function $f^>(\w)$ is plotted for both choices in the top panels
of \figsref{figLor} and \ref{figGauss} for $\eta=0.2$.

Based on the same argument as is the basis for \secref{secTimeEvol}
in this paper (approximate equivalence of Fourier and Chebyshev
expansion), \oc{wolf14} showed the decrease of 
Chebyshev moments for superpositions of Lorentzians 
and Gaussians, to be approximately exponential  
and $\propto e^{-\alpha n^2}$, respectively. 
This behavior is observed in both center panels 
of \figsref{figLor} and \ref{figGauss}. For high values
of $n$, in \figref{figGauss}, the decrease $\propto e^{-\alpha n^2}$
transitions into an exponential decrease, which is not in 
contradiction with the result of \oc{wolf14}. 
This, as the discussion of \secref{secTimeEvol}, make 
statements for intermediate values of $n$.

In the bottom panels of \figsref{figLor} and \ref{figGauss}
we then show the error obtained for the different methods of reconstruction.
Linearly predicting $\wt A^>(\w)$ yields
considerably lower errors than linearly
predicting the ``full'' 
spectral function $A(\w)$. Only in the case of Lorentzians
(lower panel of \figref{figLor}), 
using $A(\w)$ leads to lower errors for values of $n$.
\begin{figure}
\ig{\numA}{pdf}{\figdirCheb fig_gauss_rho}
\ig{\numA}{pdf}{\figdirCheb fig_gauss_mu}
\ig{\numA}{pdf}{\figdirCheb fig_gauss_err_cmpTL}
\caption{(Color online) 
\tbf{Top}: Test function consisting of two Gaussians \eqref{testf}. 
\tbf{Center}: Corresponding Chebyshev moments
in the three different setups $\mu^>$, $\wt\mu^>$ and $\mu$,
analogously to the bottom panel of \figref{figIntegrand}.
\tbf{Bottom}: Error of reconstructed spectral function,
analogously to the bottom panel of \figref{fig_SIAMcompareSetups}.
All of this is for $a=100$.
}
\label{figGauss}
\end{figure}

\begin{figure}
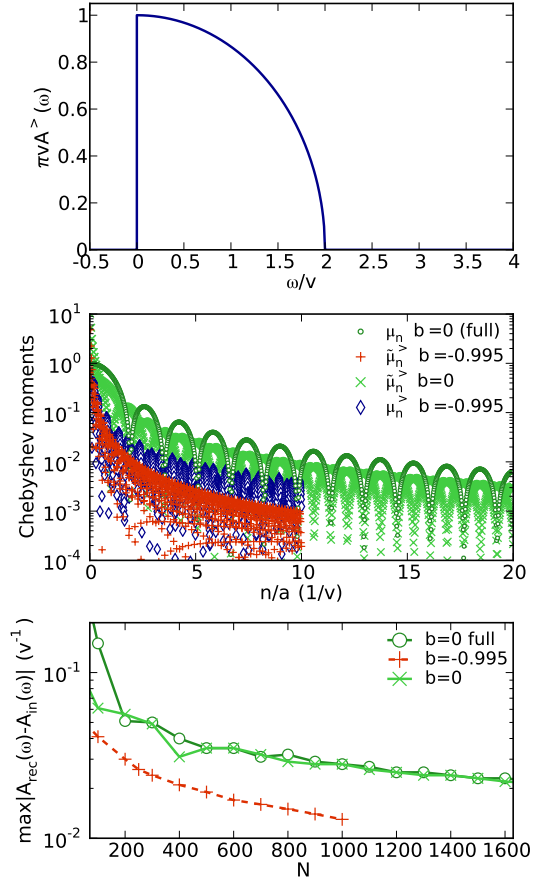

\ig{\numA}{pdf}{\figdirCheb fig_U0_rho}
\ig{\numA}{pdf}{\figdirCheb fig_U0_mu}
\ig{\numA}{pdf}{\figdirCheb fig_U0_err_cmpTL}
\caption{(Color online) 
\tbf{Top}: Particle spectral function of 
half-filled non-interacting SIAM. 
\tbf{Center}: Corresponding Chebyshev moments
in the three different setups $\mu^>$, $\wt\mu^>$ and $\mu$,
analogously to the bottom panel of \figref{figIntegrand}.
\tbf{Bottom}: Error of reconstructed spectral function,
analogously to the bottom panel of \figref{fig_SIAMcompareSetups}.
}
\label{figU0}
\end{figure}

Finally, in \figref{figU0}, we show results for 
the spectral function of the half-filled non-interacting SIAM
with semi-elliptic bath density of states,
which itself is semi-elliptic,
\eq{
A^>(\w) =
\left\{
\begin{array}{ll}
\qquad 0 & \tx{for }  \w < 0, \\
\qquad 0 & \tx{for } \w > 2v, \\
\frac{1}{2v\pi} \sqrt{4 - \Big(\frac{\w}{v}\Big)^2} & \tx{else.}
\end{array}
\right.
}
$A^>(\w)$ has a kink at $\w=2v$, as can be
seen in the top panel of \figref{figU0}.
Therefore, Chebyshev moments decrease only algebraically,
as seen in the center panel of \figref{figU0}.

If we consider the error of the linear-prediction-based 
reconstructed $A^>(\w)$, shown in the bottom panel of \figref{figU0}, 
we see that this yields much better results than the estimate  \eqref{convEstim}
gives for a plain truncation of an algebraically decreasing series. 
Concerning the comparison between the two methods of 
reconstructing $\wt A^>(\w)$ and the full $A(\w)$, we observe that 
$\wt A^>(\w)$ yields to smaller errors throughout.

We finally note that studying the spectral function of the non-interacting
SIAM with a constant bath density of states as in \oc{holzner11} and \oc{ganahl14}, 
does not constitute a more general case.
The analytic expression for this is very 
close to a single Lorentzian.

\section{Linear prediction}
\label{secLP}

In the context of time evolution \tit{linear prediction}
has been long established in the DMRG community,\cp{white08,barthel09}
but it has only recently been applied to the computation of Chebyshev moments.\cp{ganahl14,wolf14}
The optimization problem for the sequence $\mu_n$ becomes linear,
if the sequence can be defined \tit{recursively}
\eq{
\tilde{\mu}_n = - \sum_{i=1}^p a_i \mu_{n-i},   \label{eqLinPredAnsatz}
}
which is easily found to be equivalent to \eqref{eqLPansatz}.\cp{barthel09}
The strategy is then as follows. Compute $N_c$ Chebyshev moments, and predict 
moments for higher values of $n$ using \eqref{eqLinPredAnsatz}.

The coefficients $a_i$ are optimized by minimizing the least-square error
$\sum_{n\in \mcal N_\tx{fit} } |\tilde \mu_n-\mu_n|^2$
for a subset $\mcal N_\tx{fit}=\{N_c-n\tl{fit},\dots,N_c-1,N_c\}$ of the computed data.
We confirmed $n\tl{fit}=N_c/4$ 
to be a robust choice, (i) small enough
to go beyond complicated low-order
(short-time) behavior and (ii) 
large enough to have a good statistics for the fit. 
Earlier,\cp{wolf14} we chose $n\tl{fit}=N_c/2$, which leads to a better
statistics for the fit. But this improvement is not important, as 
we do not deal with stochastic data.
Minimization yields
\eq{
R \bm{a} & = -\bm{r}, \quad \bm{a}=-R^{-1}\bm{r}, \\ 
R_{ji} & = \sum_{n\in\mcal N_\tx{fit}} \mu^*_{n-j} \mu_{n-i},\quad
r_{j} = \sum_{n\in\mcal N_\tx{fit}} \mu^*_{n-j} \mu_n. \non
}
Linear prediction is more prone to \tit{overfitting} 
if choosing $p$ to be very high. Therefore one should
restrict the number of coefficients to $p=\min(n\tl{fit}/2,100)$.
Furthermore, one adds a small constant $\delta=10^{-6}$ to 
the diagonal of $R$ in order to enable the inversion of the 
singular matrix $R$.
Defining\cp{barthel09}
\[M=\left(\begin{array}{ccccc}
-a_1 & -a_2 & -a_3 & \dots & -a_{p}\\
1   & 0     &   0  & \dots & 0\\
0   & 1     &   0  & \dots & 0\\
\vdots   & \ddots  & \ddots &\ddots & \vdots\\
0  & 0       & \dots &1 & 0\\
\end{array}\right),
\]
one obtains the predicted moments 
$\tilde\mu_{N_c+n} = (M^n \bm{\mu}_{N_c})$, where 
$\bm{\mu}_{N_c} = (\mu_{N_c-1}\; \mu_{N_c-2}\; \dots\; \mu_{N_c-p})^T$. 
The matrix $M$ might have eigenvalues with absolute value larger than $1$,
either due to numerical inaccuracies or due to the fact that linear prediction
cannot be applied as $\mu_n$ rather increases than decreases on the 
training subset $\mcal N_\tx{fit}$.
In order to obtain a convergent prediction, we set the weights 
that correspond to these eigenvalues to zero measuring the ratio of
the associated discarded weight compared to the 
total weight. If this ratio is higher than a few percent, 
we conclude that linear prediction cannot be applied. One can 
then restart the computation to increase the number of computed moments $N_c$,
and try applying linear prediction 
for a higher number of moments.

\section{Discretization for odd $L$}
\label{secDiscr}

In the case of odd $L$ we cannot equate the central interval boundary 
with 0 as in \eqref{wL2}. Instead we have to choose the correct width 
for the ``central interval" by choosing the neighboring boundaries
$\w_{(L+1)/2}$ and $\w_{(L+1)/2+1}$
correctly. This is achieved by subtracting an offset $\Delta w$ from each 
of the positive boundaries defined in \eqref{cosDiscr}, such that for $l = 1,2,...$
\eq{
\w_{(L+1)/2+l} & = a\big(\cos(\theta_{(L+1)/2+1} + l \Delta\theta)-b - \Delta\w\big), \non \\
\Delta\theta &= \tfrac{2}{L} (\theta_{L+1}-\theta_{(L+1)/2+1}),  \non\\
\theta_{(L+1)/2+1} & = \arccos b , \non\\
\theta_{L+1} & = \arccos(b + \tfrac{\w_{L+1}}{a}),    \label{cosDiscrOdd} \\
\Delta\w & = \frac{1}{2}\big(\cos(\theta_{(L+1)/2+1} + \Delta\theta)-b\big). \non 
}
For negative boundaries $\Delta\w$ has to be added instead of subtracted.

\section{Time evolution by Fourier transform}
\label{secStanTimeEvol}

Given the Chebyshev expansion in frequency space
\eq{
A^>(\w) & = \frac{1}{a} \mathcal{A}^>\big(\frac{\w}{a}+b\big), \non\\
\mathcal{A}^>(x) & = \sum_n w_n(x) \mu_n T_n(x),
}
we can obtain the time 
evolution of a \tit{single} Green function,
but not for the whole system state, by Fourier transforming
\eq{
G^>(t) & = \int_{-\infty}^\infty d\w\, A^>(\w) e^{i\w t} \non\\
		   & = \frac{1}{a} \int_{-\infty}^\infty d\w \,  \mathcal{A}^>\big(\frac{\w}{a}+b\big) e^{i\w t} \non\\
	       & = \int_{-\infty}^\infty dx\, \mathcal{A}^>(x) e^{ia(x-b)t} \non\\
   	       & = \int_{-\infty}^\infty dx\, \sum_n w_n(x) \mu_n T_n(x) e^{ia(x-b)t} \non\\
   	       & = \sum_n \mu_n \int_{-\infty}^\infty dx\,  w_n(x) T_n(x) e^{ia(x-b)t} \non\\	       
	       & = e^{-iabt} \sum_n (2-\d_{n0}) (-i)^n \mu_n  J_n(at),  \label{stanTimeEvol}
}
where the last step (interchanging sum and integral) 
is only possible if the sum is absolutely convergent, or finite. 
The Fourier transform can be looked up in a handbook on integrals
\eq{
 \int_{-\infty}^\infty dx\, & w_n(x) T_n(x) e^{ia(x-b)t}   \non \\
 & = (2-\d_{n0}) e^{-iabt} \int_{-\infty}^\infty dx\,  \frac{T_n(x) e^{iaxt}}{\pi \sqrt{1-x^2}}  \non\\
 & = (-i)^n (2-\d_{n0}) e^{-iabt}  J_n(at).
}

\section{Comparison to time evolution}
\subsection{Standard recursion for cosine}
\label{seqStandardRec}

As usual for a vector space of orthogonal polynomials,  the space of cosine functions 
$\{\cos(nx)\}$, where $n\in\mathbb{N},~x\in\mathbb{R}$, can be generated using a three-term recursion formula
\eq{
\cos(nx) = 2\cos(x) \cos((n-1)x) - \cos((n-2)x),
}
which can be proven using addition theorems. 

Rewriting this in the operator-valued form for the argument $\mcal{H}$,
and acting on $\ket{\psi_0}$ yields for the definition of  $\ket{\phi\tl{cos}(t_n)}$ in \Eqref{eqEulerDiscr}
\eq{
\ket{\phi\tl{cos}(t_n)} 
= 2\cos \mathcal{H} \ket{\phi\tl{cos}(t_{n-1})}  - \ket{\phi\tl{cos}(t_{n-2})}.
}

But this provides no solution for our problem
as the action of  $\cos\mcal{H}$ 
on $\ket{\phi\tl{cos}(t_{n-1})}$ is not known.

\subsection{Shifted cosine}
\label{secShiftedCos}

Using an addition theorem
\eq{   \label{shiftCos}
\cos\big(n & (\tfrac{\pi}{2} - \a )\big)  = \\
= & \cos\big(n\tfrac{\pi}{2}\big) \cos(n \a) 
+ \sin\big(n\tfrac{\pi}{2}\big) \sin(n \a ) \non\\
= & 
\left\{
\begin{array}{cl}
\cos(n \a ) &  \tx{ if } \frac{n}{4} \in \mathbb{N} \\
\sin(n \a ) & \tx{ if } \frac{n+2}{4} \in \mathbb{N} \\
\frac{1}{\sqrt{2}} \big( \cos(n \a ) + \sin(n \a ) \big) & \tx{ else.} \\
\end{array}
\right. \non
}
\Eqref{eqApprox} follows if setting $\a= \mcal{H}$.

\subsection{Sine term}
\label{secSine}

Changing the initial conditions of \eqref{eqRecTn} 
generates the polynomials $T_n'=\sin(n\arccos(x))$
\eqs{
T_n'(x) = 2x T_{n-1}'(x) - T_{n-2}'(x),  \label{eqRecTn}\\
T_1'(x) = \sqrt{1-x^2}, \quad T_0'(x) = 0,
}
which approximates the sine function, in the same way as \eqref{eqApprox}
approximates the cosine
\eq{
T_n'(\mcal{H}) \simeq  \sin\Big(n \mcal{H}\Big) & \tx{ if } n/4 \in \mathbb{N} .
}

\subsection{Bound for Arccos}
\label{secBoundArccos}

Bounding the Arccos works as follows
\eq{
\arccos(x) 
& = \frac{\pi}{2} - \sum_{n=0}^\infty \frac{ \binom{2n}{n} x^{2n+1}}{4^n (2n+1)} \non\\
& = \frac{\pi}{2} - x - r(x) \non\\
r(x) & = x^3 \Big(\frac{1}{6}  + \Big\vert \sum_{n=2}^\infty \frac{ \binom{2n}{n} x^{2n-2}}{4^n (2n+1)} \Big\vert \Big).
}
Using $\arcsin(1) = \sum_{n=0}^\infty \frac{ \binom{2n}{n}}{4^n (2n+1)} = \frac{\pi}{2}$ we can bound
\eq{
\abs{r(x)} 
& =  |x^3| \Big(\frac{1}{6}  + \Big\vert \sum_{n=2}^\infty \frac{ \binom{2n}{n} x^{2n-2}}{4^n (2n+1)} \Big\vert \Big) \non\\
& < |x^3| \Big(\frac{1}{6}  + \Big\vert \sum_{n=2}^\infty \frac{ \binom{2n}{n} }{4^n (2n+1)} \Big\vert \Big)  \non\\
& = |x^3| \Big(\frac{1}{6}  + \frac{\pi}{2} - 1 - \frac{1}{6} \Big)   \non\\
& = |x^3| \Big(\frac{\pi}{2} - 1 \Big) \non\\
& < \frac{2}{3} |x^3| .
}

\subsection{Error computation}
\label{secError}

The approximation in \eqref{eqApprox} is based on the Taylor expansion
\eq{  \label{eqTaylor}
\arccos(\mcal{H}) & = \frac{\pi}{2} - \mcal{H} + \frac{1}{6} \mcal{H}^3 + \mathcal{O}(\mcal{H}^5), \\
\mcal{H} & = \frac{H-E\tl{ref}}{a}, \non
}
which reflects the fact that the arcus cosine is well 
approximated already by the leading linear term around $x=0$. 

The approximation of $\ket{\phi(t_n)}$ that has 
been generated in this way, is good if $a$ is large enough
and becomes exact for $a\ra\infty$. But how large does one 
have to choose $a$ in practice
in order for $\e(t_{n'})$ to be bounded by the wished accuracy?

Consider the decomposition of the initial state 
in eigen states $\ket{E_n}$ of $\mathcal{H}$
\eq{
\ket{\psi_0} = \sum _k c_{k}  \ket{E_k},
} 
and, defining $\w_k = \frac{E_k-E\tl{ref}}{a}$,
therefore 
\eq{
T_n (\mcal{H}) \ket{\psi_0} = \sum _k c_{k}  T_n (\w_k) \ket{E_k}.
}
We are now only interested in indices $n'$ that are multiples of $4n$, 
as only those have the interpretation of a time-evolved state. Therefore, 
\eq{  
T_{n'} (\w_k)  
= &  \cos(n' \arccos(\w_k))      \non \\
= &  \cos(n' \w_k - n' r_k)  \qquad \tx{ as } n'=4n,\,\tx{see Eq.}\,\eqref{shiftCos}     \non  \\
= &  \cos(n' \w_k) \cos(n' r_k)  + \sin(n'\w_k) \sin(mr_k)   \non  \\
 = & \cos(n' \w_k) + \e_k(n')  , \non  \\  
 \e_k(n') & =  \cos(n' \w_k) (\cos(mr_k) - 1) \non\\
 &\qquad+  \sin(n'\w_k)\sin(mr_k)    \non
}
Up to here everything was exact.

Now we can strictly bound the absolute value of the error term
using $|\sin(r_k)| < |r_k|$ and $|\cos(r_k)-1|< \frac{r_k^2}{2}$
and trivially bounding $\cos(n' \w_k)$ and $\sin(n' \w_k)$ by one,
\eq{
|\e_k(n')|  <  \tfrac{1}{2} n'^2 r_k^2 +  n' |r_k|
}
Let us now define the energy eigenvector $\ket{E\tl{max}}$ for which the error $\e_k$ 
becomes maximal, which is the one for which $\w\tl{max}$ is maximal, \ie
\eq{
\w\tl{max} = \max_{\ket{E_k} \in \ket{\psi_0}} \w_k
}
with which we compute $r\tl{max}$ and $\e\tl{max}$. We can then  
simplify further
\eq{
 \sum _k c_{k}  \e_k(n') \ket{E_k} 
 <  \e\tl{max}(n') \sum _k c_{k} \ket{E_k}   
 = \e\tl{max}(n') \ket{\psi_0}.   \non
}
We therefore arrive at
\eq{
T_{n'} (\mcal{H}) \ket{\psi_0} 
= \cos(n' \mcal{H}) \ket{\psi_0}
+ \e(n') \ket{\psi_0}  
}
The value of $\w\tl{max}$ is determined by the cutoff of the distribution
of eigenvectors $\ket{E_k}$ in $\ket{\psi_0}$. This can be a strict cutoff
or a few standard deviations of Gaussian distribution, 
beyond which no contributions with numerically
measurable weight occur. Let denote this cutoff or width $\sigma$ and 
define it analogously to $\w\tl{max}$, \ie
\eq{
\sigma = \max_{\ket{E_k} \in \ket{\psi_0}} |E_k-E\tl{ref}| \quad \Rightarrow \quad \w\tl{max} = \frac{\sigma}{a}
}
If, e.g., $\ket{\psi_0}$ is constructed by applying a single-particle operator 
to an eigenstate (\eg the ground state) of $H$, $\sigma$ is the single particle
bandwidth $W\tl{single}$ times a small factor of order 1.

Finally we need to bound the error term $r_k$ (\appref{secBoundArccos})
\eq{ \label{rk}
r_k = \frac{1}{6} \w_k^3 + \mcal{O}(\w_k^5) , \quad \abs{r_k} <  \frac{2}{3} |\w_k^3|.
}
Using the definition of $\sigma$, let us now bound $\e(n')$
\eq{
|\e(n')|
& <   |\e\tl{max}(n')|       \non\\
& <   \tfrac{1}{2} n'^2 r\tl{max}^2 +  n' |r\tl{max}|   \non\\
& <   \tfrac{1}{2} n'^2  (\tfrac{2}{3})^2 \w\tl{max}^6 +  n'  \frac{2}{3} |\w\tl{max}^3|    \non\\
& =   \tfrac{1}{2} n'^2  (\tfrac{2}{3})^2 \Big(\frac{\sigma}{a}\Big)^6 +  n'  \frac{2}{3} \Big(\frac{\sigma}{a}\Big)^3 \non \\
& <   \frac{3}{2} n' \frac{2}{3} \Big(\frac{\sigma}{a}\Big)^3 \qquad \tx{if} \quad  n' < n'\tl{ref} \non \\
& = \frac{n'}{n'\tl{ref}}, \quad n'\tl{ref} = \Big(\frac{a}{\sigma}\Big)^3.  \non
}

Or expressing this in units of time
\eq{
|\e(t_{n'})|
& =  \frac{t}{t\tl{err}} \qquad \tx{if} \quad  t < t\tl{err}  \non \\
t\tl{err} & = \frac{a^2}{\sigma^3} \label{errorTimeApp}
}

Inserting typical values, where $v$ is a hopping energy
for a single particle process: $\sigma=2v$, $a=100v$ one finds 
$t\tl{err} = \frac{1250}{v}$. 
The accumulated error $\e(t_{n'})$ therefore remains smaller 
than $10^{-2}$ if $t<12.5\frac{1}{v}$.

% \bibliography{lit}

\end{document}